\documentclass[prd,aps,reprint,onecolumn,superscriptaddress,tightenlines,nofootinbib,eqsecnum,preprintnumbers,longbibliography,12pt]{revtex4-2}
\usepackage{epsfig,amsfonts,mathrsfs,amsmath,amssymb,graphicx,color,slashed,multirow}
\usepackage{amsmath,latexsym,amssymb,graphicx,slashed,hyperref,color,enumerate,url,etoolbox,cancel,gensymb,physics}
\hypersetup{colorlinks,citecolor= nicegreen,linkcolor= nicered}
\usepackage[usenames,dvipsnames]{xcolor}
\definecolor{nicered}{rgb}{0.7,0.1,0.1}
\definecolor{nicegreen}{rgb}{0.1,0.5,0.1}
\definecolor{niceblue}{rgb}{0.1,0.2,0.6}

\usepackage{cancel}
\usepackage{bbm}
\usepackage{physics}
\usepackage{dsfont}
\usepackage{cancel}
\definecolor{DarkerRed}{rgb}{0.666,0,0}
\definecolor{DarkerGreen}{rgb}{0,0.4,0 }
\definecolor{DarkerBlue}{rgb}{0,0,0.4 }
\definecolor{DarkerPurple}{rgb}{0.45,0.1,0.45 }
\definecolor{DarkerOrange}{rgb}{0.66,0.35,0.3 }

\usepackage[normalem]{ulem} 

\begin{document}

\def\Carleton{Ottawa-Carleton Institute for Physics, Carleton University, Ottawa, ON K1S 5B6, Canada}
\def\TRIUMF{TRIUMF, 4004 Wesbrook Mall, Vancouver, BC V6T 2A3, Canada}
\def\UW{Department of Physics, University of Wisconsin-Madison, Madison, WI 53706, USA}
\def\ANL{HEP Division, Argonne National Laboratory, Argonne, IL 60439, USA}

\title{
Novel Phases of a Baryon-Dense QCD-like Theory
}

\author{Yang Bai}
\email{yangbai@physics.wisc.edu}
\affiliation{\UW}
\affiliation{\ANL}
\author{Carlos Henrique de Lima}
\email{cdelima@triumf.ca}
\affiliation{\TRIUMF}
\author{Daniel Stolarski}
\email{stolar@physics.carleton.ca}
\affiliation{\Carleton}


\begin{abstract}
We investigate the phases of a strongly coupled QCD-like theory at finite baryon chemical potential using s-confining supersymmetric QCD deformed by anomaly-mediated supersymmetry breaking. Focusing on the case of three colors and four flavors, we identify novel phases including spontaneous breaking of baryon number and/or parity.  Both first-order and second-order phase transitions are observed as the baryon chemical potential is varied. These findings may offer insights into possible phases of real QCD at intermediate baryon densities.  
\end{abstract}


\maketitle
\section{Introduction}
Understanding quantum chromodynamics (QCD) at high baryon densities is a longstanding goal of high energy physics~\cite{Rajagopal:2000wf,Alford:2001dt}. The inherent strong dynamics of QCD make the analysis of the phase diagram only controlled in the high-energy regime and consequently high densities. In this regime, a plethora of symmetry breaking patterns appear, depending on the number of flavors, such as color superconductivity, color-flavor locking, and dynamical breaking of baryon number~\cite{Schafer:2000tw,Alford:2000ze}. At lower densities, extracting the phase diagram is significantly more challenging and usually relies on lattice~\cite{Vovchenko:2018zgt,DElia:2002tig,deForcrand:2006ec} or phenomenological models~\cite{Baym:2017whm,Wu:2020qgr}. More importantly, obtaining results with massless fermions is intrinsically challenging for these approaches. In this work, we build upon recent results of deforming Supersymmetric QCD (SQCD) with anomaly mediated supersymmetry breaking (AMSB) to obtain the phase diagram of QCD with four massless flavors for energies below the confinement scale.

SQCD provides a calculable framework to study strongly coupled QCD-like theories. Exact results in SQCD  elucidate low-energy descriptions for arbitrary numbers of colors, $N_c$, and fundamental flavors, $N_f$, making it a powerful tool for exploring non-perturbative results in strongly coupled quantum field theory~\cite{Seiberg:1994bz,Seiberg:1994pq}. The theory with $N_f=N_c+1$ is ``s-confining'', meaning that the low energy theory can be described in terms of gauge invariant composites of baryons and mesons, exhibiting confinement but without chiral symmetry breaking at the origin of the moduli space. This makes it particularly closely related to its non-supersymmetric counterpart, QCD, which is also described at low energy by baryons and mesons. 

Coupling SQCD to a non-zero baryon chemical potential $\mu_B$ makes the theory unstable~\cite{Harnik:2003ke}. The finite baryon density induces a tachyonic direction in the scalar potential, causing the theory to flow to a non-perturbative regime. This can be stabilized by adding masses to the theory, either a supersymmetric mass or a soft supersymmetry (SUSY) breaking mass. Even with those masses, the theory is still unstable for sufficiently large $\mu_B$. 

In this work, we explore SQCD coupled to anomaly mediated SUSY breaking~\cite{Randall:1998uk,Giudice:1998xp,Pomarol:1999ie,Jung:2009dg} with non-zero baryon chemical potential. Anomaly mediation is extremely useful in analyzing strongly coupled quantum field theories as it is UV insensitive~\cite {Murayama:2021xfj}, so the soft terms can be calculated precisely in the low energy effective theory with no knowledge of how to map any operator from the UV to the IR. Recent work has explored the dynamics of several SQCD theories coupled to AMSB~\cite{Murayama:2021xfj,Csaki:2021aqv,Csaki:2021jax,Bai:2021tgl,Luzio:2022ccn,Kondo:2022lvu,Csaki:2022cyg,deLima:2023ebw,Csaki:2024lvk,Leedom:2025mcg,Goh:2025oes,Kondo:2025njf}, including the s-confining regime. 
We focus specifically on s-confining SQCD coupled to anomaly mediation with a non-zero baryon chemical potential $\mu_B$. This model, particularly with the choice of three colors, closely resembles QCD with four massless quarks, where the additional states decouple as the SUSY breaking parameter is made large. 

The instabilities of the supersymmetric theory~(and also with small SUSY breaking) at finite chemical potential were initially identified in~\cite{Harnik:2003ke} using a different soft SUSY breaking.\footnote{In principle, any results obtained from deforming SUSY to non-SUSY do not depend on the SUSY breaking scheme. However, since in most cases, the SUSY breaking parameter is not strictly infinite, each SUSY breaking approaches the non-SUSY regime differently. At the same time, some methods are more calculable than others for strongly coupled theories.} In~\cite{Harnik:2003ke}, it was assumed that the symmetry-breaking pattern would remain the same as the given runaway direction. However, this is not necessarily the case since the vacuum becomes dominated by strong dynamics, which are incalculable even with the SUSY machinery.

In this work, we highlight a stabilization mechanism that allows us to extract the phase diagram of the QCD-like theory. The stabilization comes from the interplay between strong dynamics and SUSY breaking. In the low energy hadronic phase, we adiabatically turn on a finite baryon chemical potential. The nonzero baryon chemical potential eventually creates a runaway instability, pushing the sbaryon vacuum expectation value (VEV) towards high energy where the theory is described by a quark and gluon phase. When the sbaryon VEV (the order parameter of this transition) starts to approach the confinement scale, higher order terms in the Kähler potential induced by the strong dynamics become important. We show how these operators can stabilize the theory below the confinement scale, keeping the theory in the hadron phase under perturbative control.

With the theory stabilized, we can explore the vacuum structure as a function of the baryon chemical potential $\mu_B$ as well as the size of the SUSY breaking. Because SUSY breaking is communicated via anomaly mediation, once the anomaly mediation scale $m_{3/2}$ is specified, all other SUSY breaking soft terms are calculable. There are still low energy parameters in the supersymmetric theory that are not predicted, and we vary their values using a few different schemes. We find several possible vacua for this theory, some of which, as far as we know, are novel for the QCD literature. The Lagrangian of the supersymmetric theory contains a $SU(4)_L\times SU(4)\times U(1)_B\times U(1)_R$ global symmetry. When coupled to AMSB, the $U(1)_R$ is explicitly broken. The vacua we identify contain the following remnant global symmetries:
\begin{itemize}
    \item {\color{DarkerRed}$SU(4)_{L} \times SU(4)_{R} \times U(1)_{B}$}\,,
    \item {\color{DarkerGreen}$SU(4)_{V} \times U(1)_{B}$}\,,
    \item {\color{DarkerPurple}$SU(3)_V\times U(1)_{ \rm res}$}\,,
    \item {\color{DarkerBlue}$SU(3)_{L/R}\times SU(4)_{R/L} \times U(1)_{\rm res}$}\,, 
    \item {\color{DarkerOrange}$SU(3)_L \times SU(3)_R$}\,. 
\end{itemize}
The first two are the standard s-confining and QCD-like vacua, respectively. The others include vacua with spontaneous breaking of baryon number and/or parity, or a preserved left-right symmetry. The last vacuum can exhibit either parity preservation or spontaneous parity breaking, depending on the specific low energy constants. We also explore the order of phase transitions between these vacua, noting the presence of both first and second-order phase transitions depending on the specific value of the low energy constants. 

As we can access the dynamics of the hadronic phase in a controllable way,  the symmetry breaking patterns highlighted in this work could be explored in an analogous nuclear system. The more immediate parallel would be the spontaneous breaking of Isospin~\cite{Palkanoglou:2024epd,Palkanoglou:2025guj,Frauendorf:2014mja,Dean:2002zx,Sedrakian:2018ydt,Bertsch:2009xz}, as it occurs in heavy nuclei. But there could be more complex patterns that are suppressed by the strange and charm quark masses, softly breaking the patterns shown in this work. 

To construct the complete phase diagram of this QCD-like theory, it is necessary to also include temperature effects. The inclusion of temperature is intrinsically more challenging as no spurion-like method exists to keep SUSY controllable. We explore the inclusion of temperature and the combination with the results of this work in~\cite{tempAMSB}.

Beyond offering insight into QCD, the approach developed in this work can be generalized to other theories with massless matter at nonzero chemical potential. Such theories are generically difficult to study using lattice techniques or other theoretical tools. This provides a calculable framework to explore more generic strongly coupled theories at finite densities and can be used, for example, in the exploration of strongly coupled dark sectors. 

The organization of this paper is as follows. In Section~\ref{sec:prelim}, we review results from the literature on SQCD, AMSB, and finite baryon chemical potential that we use in our analysis. In Section~\ref{sec:analysisMU}, we describe the runaway directions in the high and low energy phases of the theory and show how, in the hadron phase, they can be stabilized by higher order Kähler terms. In Section~\ref{sec:cons}, we compute the scalar potential of the theory and describe the different phases. In Section~\ref{sec:phase}, we perform a numerical exploration of the phase structure of the theory and show phase diagrams for different choices of low energy constants. In Section~\ref{sec:transitions}, we explore a few possible phase boundaries and the nature of the phase transitions. We conclude in Section~\ref{sec:conc}, and various technical details are given in the appendices.

\section{Preliminaries}
\label{sec:prelim}

In this section, we review standard results from the literature that we use in our analysis. We describe anomaly-mediated SUSY breaking, s-confining SQCD, and the inclusion of baryon chemical potential in SUSY models.

\subsection{Anomaly mediation}
Anomaly-mediated SUSY breaking (AMSB) is one of the simplest and most controlled mechanisms of SUSY breaking~\cite{Pomarol:1999ie,Randall:1998uk,Giudice:1998xp,Jung:2009dg}. AMSB has the unique feature that connects the SUSY breaking to the conformal violation of the theory, giving more theoretical control over SUSY-breaking effects at low energies in confining theories~\cite{Murayama:2021xfj,Csaki:2022cyg,deLima:2023ebw}. The SUSY breaking effects are fully calculable in terms of one SUSY breaking scale $m_{3/2}$. The SUSY breaking parameter is generally complex, but its phase can be rotated away to the vacuum angle of the gauge theory~\cite{Csaki:2024lvk}. We therefore take $m_{3/2}$ to be real for the remainder of this paper. At the tree level, the SUSY-breaking scalar potential for a general $N=1$ supersymmetric theory is given by
\begin{align} \label{eq:treeB}
V_{\text {\bcancel{tree}}} & =m_{3/2}\left(\partial_i W g^{i j^*} \partial_j^* K -3\,W\right)+ \text{h.c.} +|m_{3/2}|^{2}\left(\partial_i K g^{i j^*} \partial_j^* K-K\right)  \, ,
\end{align}
where $W(\varphi_i)$ is the superpotential of the superfields $\varphi_i$, $K$ is the Kähler potential, $g_{i j^*}=\partial_{i}\partial_j^*K$ is the Kähler metric, and $g^{i j^*}$ its inverse. For a canonical Kähler potential, the tree-level SUSY-breaking potential reduces to
\begin{align}
V_{\text {\bcancel{tree} }}=m_{3/2}\left(\varphi_i \frac{\partial W}{\partial \varphi_i}-3\, W\right) + \text{ h.c.} \, .
\end{align}
Since anomaly mediation is tied to the breaking of conformal symmetry, there are also loop-level SUSY breaking effects from the superconformal anomaly. These include trilinear couplings, scalar masses, and gaugino masses:
\begin{align}
A_{i j k} & =-\frac{1}{2}\,\left(\gamma_i+\gamma_j+\gamma_k\right) m_{3/2} \, , \\
\label{eq:softmass} m_i^2 & =-\frac{1}{4}\, \dot{\gamma}_i\, |m_{3/2}|^2 \, , \\ 
\label{eq:gauginomass} m_\lambda & =-\frac{\beta\left(g^2\right)}{2 g^2} m_{3/2} \, .
\end{align}
Here, $g$ is the gauge coupling, $\beta\left(g^2\right)$ is the beta-function of the gauge coupling running, and $\gamma_i$ is the anomalous dimension of the chiral superfields.\footnote{We define the anomalous dimension, its derivative, and the $\beta$ function as $\gamma_i=\mu \frac{d}{d \mu} \ln Z_i, \dot{\gamma}=\mu \frac{d}{d \mu} \gamma_i$, and $\beta\left(g^2\right)=\mu \frac{d}{d \mu} g^2$. }

\subsection{s-confining SQCD with $N_c\geq 3$}

In this work, we study s-confining SQCD, which has an $SU(N_{c})$ gauge theory with $N_{f}=N_{c}+1$ vectorlike fundamental flavors~\cite{Seiberg:1994pq,Murayama:2021xfj,Csaki:2022cyg,deLima:2023ebw}. In this section, we work out the results for arbitrary $N_c$. In Section~\ref{sec:cons} when we perform numerical analysis, we specialize to $N_{c}=3$ and $N_f=4$.

In the UV, the theory is described by quarks $q_{\alpha i}$ and anti-quarks $\bar{q}^{\alpha i}$, where we use Greek letters for gauge indices and Latin letters for flavor indices. The theory has an $SU(N_f)_L\times SU(N_f)_R\times U(1)_B\times U(1)_R$ global symmetry that forbids a superpotential.  Flowing to the IR, the theory confines at a scale $\Lambda$. For $N_c \geq 3$, the low energy is described by gauge invariant composites~\cite{Seiberg:1994pq}
\begin{eqnarray} 
B^i &=& \varepsilon^{\alpha_1 ... \alpha_{N_c}} \varepsilon^{i_1 ... i_{N_c}i} \, q_{\alpha_1 i_1}\, ... \, q_{\alpha_{N_c} i_{N_c}} \,, \nonumber\\
\widetilde{B}_i &=& \varepsilon_{\alpha_1 ... \alpha_{N_c}} \varepsilon_{i_1 ... i_{N_c}i} \, \bar{q}^{\, \alpha_1 i_1}\, ... \, \bar{q}^{\, \alpha_{N_c} i_{N_c}} \,, \nonumber\\
M^i_j &=& q_{\alpha j} \, \bar{q}^{\, \alpha i} \, .
\end{eqnarray}
Like real QCD, the theory is invariant under the discrete charge conjugation (C) and parity (P) symmetries. In the supersymmetric theory, these symmetries are preserved at low energies. The strong dynamics induce the following dynamical superpotential for the baryon and meson superfields at low energies 
\begin{align}
    W = \frac{1}{\Lambda^{2N_c-1}} \left( \det M - \widetilde{B}MB \right) \, .
\end{align}
From the non-trivial anomaly matching conditions, it is expected that the Kähler potential is regular at the origin for the meson and baryon superfields. We thus can canonically normalize the fields, such that the superpotential can be written as
\begin{align}
W=\lambda \frac{\operatorname{det} M}{\Lambda^{N_c-2}}-\kappa \widetilde{B} M B \, ,
\label{eq:superpotential}
\end{align}
with the low energy coupling constants $\kappa$ and $\lambda$ that become weakly coupled at scales well below $\Lambda$. The value of these couplings is connected to the gauge coupling or, more directly at low energy, to $\Lambda$. The exact mapping relies on non-perturbative physics, and we take here an agnostic view to explore all possible values that live in the perturbative domain. The SUSY preserving theory has several flat directions both in the UV and the IR that create a rich moduli space. The origin of the moduli space preserves chiral symmetry for all values of $\kappa$ and $\lambda$, making it an s-confining theory. The low energy supersymmetric spectrum has only meson and baryon superfields, and they are massless at the origin.

We now couple the theory to AMSB beginning in the quark and gluon phase. In this UV phase, the theory is classically conformal, so leading SUSY breaking appears at the loop level from the gauge coupling $\beta$-function and the anomalous dimensions of the quarks. The leading order gaugino and squark masses are
\begin{align}
m_\lambda & =(2N_{c}-1) \frac{g^{2}}{16\pi^{2}} m_{3/2} \, , \\
\label{eq:UVmass} m_{q , \bar{q}}^{2} &=\frac{(N_{c}^{2}-1)(2N_{c}-1)}{N_{c}} \frac{g^{4}}{64\pi^{4}} m_{3/2}^{2}\, . 
\end{align}
The scalar masses are positive in the regime of perturbativity, and thus, the fields are driven to scales below $\Lambda$ where the composite fields better describe the theory. At this point, we see the power of the UV insensitivity of AMSB, as we do not need to follow operators from the UV to the IR, and we can directly jump to the low energy description in terms of weakly interacting composite states.

At low energies, the dynamical superpotential of Eq.~\eqref{eq:superpotential} breaks conformal symmetry, and thus, using Eq.~\eqref{eq:treeB} the leading AMSB interaction is
\begin{align}
V_{\bcancel{\text{tree}}} = \left(N_{c}-2 \right)\frac{\lambda}{\Lambda^{N_{c}-2}}m_{3/2}\det M + \text{h.c.} \, .
\label{eq:treebreak}
\end{align}
For the loop-induced contributions, we restrict our analysis to leading order; for higher loop effects, see~\cite{deLima:2023ebw}. The soft masses at leading order are given by
\begin{eqnarray}
m_M^2 &=& \frac{2N_{c}+3}{256\pi^4}|\kappa|^4 |m_{3/2}|^2 \, , \\
m_{B, \widetilde{B}}^2 &=& \frac{(N_{c}+1)(2N_{c}+3)}{256\pi^4}|\kappa|^4 |m_{3/2}|^2 \, ,
\end{eqnarray}
which are of two-loop order and positive. There is also a one-loop $A$-term, that, along with the tree level term in Eq.~\eqref{eq:treebreak}, induces symmetry breaking once $m_{3/2}$ is nonzero. 

For $N_{c}\geq 3$, we can use flavor rotations to write the distinct vacuum directions as~\cite{Csaki:2022cyg}
\begin{align}
B=\left(\begin{array}{c} \label{eq:gen}
b_{1} \\
b_{2} \\
\vdots \\
b_{F}
\end{array}\right) \, , \quad \,  \widetilde{B}=\left(\begin{array}{c}
\bar{b}_{1} \\
\bar{b}_{2} \\
\vdots \\
\bar{b}_{F}
\end{array}\right) \, , \quad \,  M=\left(\begin{array}{llll}
x_{1} & & & \\
& x_{2} & & \\
& & \ddots & \\
& & & x_{F}
\end{array}\right) \, . 
\end{align}
This most general vacuum structure can be further simplified when we analyze the AMSB potential, where most of these directions increase the vacuum energy. Using this information, we can use a simplified\footnote{This simplification is not necessarily general when including the modification from the chemical potential or higher order terms in the Kähler potential. This is further explored in Section~\ref{sec:analysisMU}.} vacuum configuration as
\begin{align}
B=\left(\begin{array}{c} \label{eq:ansatz}
b \\
0 \\
\vdots \\
0
\end{array}\right) \, , \quad  \,  \widetilde{B}=\left(\begin{array}{c}
\bar{b} \\
0 \\
\vdots \\
0
\end{array}\right) \, , \quad \,  M=\left(\begin{array}{llll}
x & & & \\
& v & & \\
& & \ddots & \\
& & & v
\end{array}\right) \, . 
\end{align}

This theory was analyzed in~\cite{Murayama:2021xfj,Csaki:2022cyg} and in-depth up to three loops in Ref.~\cite{deLima:2023ebw}. From Ref.~\cite{deLima:2023ebw}, the four-flavor anomaly-mediated SQCD (ASQCD) has a chiral symmetry-breaking and baryon number preserving vacuum that is insensitive to $m_{3/2}$ in the perturbative regime. Including a baryon chemical potential, one expects new vacuum structures.

\subsection{SQCD with finite baryon chemical potential}
The analysis of SQCD with a baryon chemical potential was initially performed in Ref.~\cite{Harnik:2003ke}, where the possible phase diagram was explored, considering the inclusion of SUSY-preserving or SUSY-breaking masses. In order to couple a SUSY theory to a finite baryon chemical potential,\footnote{We define $U(1)_B$ such that a baryon has unit charge, and thus the quarks have $1/N_c$ charge.} one can introduce the effect using the spurion formalism by including a background (fictitious) $U(1)_B$ super-gauge field $V_{B}$~\cite{Harnik:2003ke}.  The baryon chemical potential effects can be described by the condensation of the gauge boson component as 
\begin{align}
\mathcal{L}_{\mu_B}= \int \dd[4]{\theta} \left( q^{\dagger}_{i}  e^{\frac{V_{B}}{N_c} } q_{i} + \bar{q}^{\dagger}_{i}  e^{-\frac{V_{B}}{N_c}} \bar{q}_{i}\right)
\end{align}
with the following VEV configuration 
\begin{align}
\langle V_{B}\rangle = \bar{\theta}\sigma^{\mu}\theta \langle A_{\mu}^{B}\rangle \, , \, \,  \langle A_{\mu}^{B}\rangle = \left(\mu_{B} ,0,0,0\right) \, .
\end{align}
The inclusion of the baryon chemical potential explicitly breaks Lorentz invariance and supersymmetry. It also breaks the charge conjugation C, but preserves parity P. For the fermionic component, one could treat $q_i$ as the left-handed field and $\bar{q}^c_i \equiv i \sigma_2 \,\bar{q}^*_i$ as the right-handed field. Under P, one has $q_i \leftrightarrow \bar{q}^c_i$. 

To maintain control
over the analysis, we focus on the regime where the chemical potential is small compared
to the confinement scale,~$\mu_B \ll \Lambda$. If supersymmetry is unbroken (or only softly broken), we can explore the effects of introducing a non-zero baryon chemical potential. Starting in the  UV (s)quark phase, the theory has no superpotential, and the supersymmetric scalar potential is only the $D$-term
\begin{equation}
V_D = \frac{g^2}{2}\left( q^\dagger T^a q - \bar{q}^\dagger T^a \bar{q} \right)^2 ~,
\end{equation}
with $T^a$ as the $SU(N_c)$ generators.
In the weak coupling regime, the scalar fields dominate the dynamics at finite chemical potential. Including a baryon chemical potential induces a  negative mass squared for both squarks
\begin{align}
m_{q , \bar{q}}^{2} = - \frac{1}{N_c^2} \mu_{B}^{2} \, .
\end{align}
This effective negative mass drives squark condensation. Because the squark quartic interaction has a $D$-flat direction with $\langle q \rangle = \langle \bar{q} \rangle$, the condensate potential is not bounded from below, and the VEV goes to infinity. We can understand this instability by looking at a physical picture of the system. The $D$-flat direction introduces a coherently attractive force between the squarks. Then, if we start with a gas of quarks and squarks, as we increase the baryon chemical potential, the coherently attractive force dominates the dynamics and causes the system to collapse to high densities. We thus have no equilibrium finite density ground state for nonzero baryon chemical potential.\footnote{In Abelian theories~\cite{Cherman:2013rla}, it is possible to lift the $D$-flat direction by including a Fayet-Iliopoulos (FI) term. In non-Abelian theories such as SQCD, FI terms are not gauge invariant.}

The s-confining SQCD theory is therefore \emph{unstable} at finite baryon chemical potential. In principle, the theory can be made stable by including a lifting mechanism for the D-flat direction. For example, it could be possible that quantum fluctuations lift this direction dynamically~\cite{PhysRevD.7.1888,PhysRevD.9.1686}. However, in the exact SUSY theory, the effective potential~\cite{Martin:2001vx,Martin:2017lqn,Martin:2024qmi,Pickering:1996he,Grisaru:1996ve,Brignole:2000kg,GrootNibbelink:2005nez,Kuzenko:2014ypa,BUCHBINDER1994665} comes only from contributions to the Kähler potential and gauge-kinetic function and thus their contribution enters in the same directions as the tree-level potential and no lifting is possible.

There is another possibility where the system could be stable. If the degrees of freedom in the confined phase experience enough repulsive force to prevent the collapse to high densities, the theory could become trapped and stabilized at low energies. This would mean that confinement screens the microscopic attractive force. The physical picture is the following: we start with the theory at low energies and in chemical equilibrium, where the degrees of freedom are the (s)hadrons. In this phase, there is no $D$-term, but there is a dynamically generated $F$-term from the superpotential in Eq.~\eqref{eq:superpotential}. If we focus on field directions with $\langle M\rangle =0$, the scalar potential has the form
\begin{align}
\left. V_F  \right\vert_{M=0} =|\kappa|^2\,|B|^2\,|\widetilde{B}|^2.
\end{align}
We then include a finite baryon chemical potential that induces tachyonic masses to the baryon and anti-baryon scalar fields given by
\begin{align}
m_{B, \widetilde{B}}^{2} = -\,\mu_B^{2} \, .
\label{eq:BaryonSoftMass}
\end{align}
Here we encounter the same problem: if $B=0$ or $\widetilde{B}=0$, the quartic vanishes. This means that the tachyonic mass again induces a runaway, driving the theory to the UV. The theory then flows to scales above the $\Lambda$, and there, it is also unstable for any value of baryon chemical potential. To have an equilibrium finite density ground state in the low-energy theory, it is necessary to lift this runaway direction before reaching the confining scale. Unlike the UV theory, the strong dynamics induce higher-order Kähler corrections that become important as the energy approaches the confinement scale. However, even with these effects, we do not see a way to stabilize the SUSY theory since Kähler corrections do not modify this runaway direction. 

We then reach the conclusion that s-confining SQCD with finite chemical potential has no equilibrium finite density ground state solution. This breakdown of the finite density solutions is a feature of the SUSY form of the potential. SUSY breaking effects can stabilize these directions, allowing us to obtain a meaningful phase diagram.

\section{ QCD with finite baryon chemical potential from ASQCD}
\label{sec:analysisMU}

We begin the analysis by considering s-confining SQCD models $(N_F=N_c+1)$ coupled to AMSB. In the UV theory, turning on a nonzero $\mu_{B}$ induces a negative mass squared for the squarks. Including the positive contribution from AMSB provides a competing effect to stabilize the origin:
\begin{align}
m_{q , \bar{q}}^{2} &=\frac{(N_{c}^{2}-1)(2N_{c}-1)}{N_{c}} \frac{g(\mu_{\rm RG})^{4}}{64\pi^{4}} m_{3/2}^{2} - \frac{1}{N_c^2}\mu_{B}^{2} \, . 
\end{align}
Note that the anomaly mediation mass is renormalization scale dependent via the running of the gauge coupling. As the theory flows to higher energies, the anomaly-mediated mass becomes smaller as the gauge theory is asymptotically free. Thus, at very high energy, the theory is unstable for any value of $\mu_B$. For sufficiently small values of $\mu_B$, the effective squark mass will be positive at intermediate energies when $g(\mu_{\rm RG})$ is perturbative but not too small. In this regime, the theory can flow toward the low energy hadronic phase and avoid a runaway instability.

We can estimate the value of the gauge coupling for which we have confidence in the perturbative calculation. Comparing the one and two-loop contributions to the $\beta$-function~\cite{Harlander:2009mn,Jack:1996vg,Pickering:2001aq}, the theory is perturbative for $g \ll g_\text{pert} = 2 \sqrt{2} \pi  \sqrt{\frac{N_c (2 N_c-1)}{N_c (N_c-1)^2+1}}$. This expression is approximately equivalent to the standard 't Hooft coupling $g_{\text{pert}} \approx 4\pi/\sqrt{N_c}$. We can then use this value of the gauge coupling as a guide to verify what would occur in the UV theory. For $g \approx g_{\text{pert}}$, if the squark effective mass is negative, then the theory is likely in an unstable phase. As the squark VEV flows further to the UV, it probes an even weaker coupling, suppressing the anomaly-mediated mass and causing the effective mass to grow further in the negative direction, accelerating the runaway. This happens for values of $\mu^2_B$ above
\begin{align}
    \left(\mu_B^{ \text{UV-crit}} \right)^2 \sim  N_c^2 m_{3/2}^2 \, .
\label{eq:mucuv}
\end{align}

In reality, the values of the gauge coupling where we trust perturbation theory are significantly less than $g_{\text{pert}}$, which means that the values for meta-stability are also expected to have $\mu_B$ smaller than $\mu_B^{ \text{UV-crit}}$. If the system is in the regime where the squark effective soft mass is positive, then the theory flows towards the low energy, while if the chemical potential is higher than this critical value $\mu_B^{ \text{UV-crit}}$, the system is unstable and will run away in the $D$-flat direction. 

When SUSY is broken, there are contributions to the one-loop Coleman–Weinberg effective potential proportional to $m_{3/2}^{2}$.
In the runaway direction, for example when $q_{1 1} = \bar{q}^{11} = Q$ the effective potential has contributions of the form $m_{3/2}^{2} Q^{2} \log(Q^{2}/\mu_{\rm RG}^{2})$. These terms can stabilize the runaway direction, but they do so at exponentially large energy scales. We do not consider this possibility any further. Another possibility that does not work for this model is to condensate a $D$-term from gauging $U(1)_R$. The restriction in this case is the presence of a $U(1)_R^3$ anomaly, which can only be canceled by the inclusion of additional fields.

If we start with $\mu_{B}$ sufficiently small, the theory flows to the IR and is described by the confined degrees of freedom. At low energies and with no baryon chemical potential, the theory has chiral symmetry breaking for perturbative values of  $m_{3/2}$. Once $\mu_{B}$ is turned on, there is a tachyonic contribution to the sbaryon mass squared [c.f.~Eq.~\eqref{eq:BaryonSoftMass}]. As $\mu_B$ increases, the effective sbaryon mass eventually becomes negative, and a vacuum that breaks baryon number forms. The important parameter that determines the existence of such a baryon-number-breaking vacuum is given by the sbaryon mass parameter from two competing contributions, which at leading order are
\begin{align}
m_{B , \widetilde{B}}^2 = \frac{(N_{c}+1)(2N_{c}+3)}{256\pi^4}|\kappa|^4 |m_{3/2}|^2 - \mu_{B}^{2} \, .
\end{align}
The condition for developing a new vacuum is when $\mu^2_B$ is larger than 
\begin{align}
(\mu_B^{\text{IR-crit}})^{2} = m_{3/2}^{2}\frac{(N_{c}+1)(2N_{c}+3)|\kappa|^{4}}{256 \pi^{4}} \, .
\label{eq:mubc}
\end{align}
We note that as long as $\kappa$ is perturbative, $\mu_B^{\text{IR-crit}} \ll \mu^{\text{UV-crit}}_B$ defined in Eq.~\eqref{eq:mucuv}, such that we can analyze the vacuum structure of the low-energy theory under perturbative control.

For larger $\mu_B$ with a negative effective sbaryon mass, one has the same problem as in the UV, where we have a runaway in both the $B=0$ and $\tilde{B}=0$ directions. The loop-induced Coleman-Weinberg effective potential is too small to lift the runaway direction before reaching the confining scale. Nevertheless, we have a different situation because of SUSY breaking effects. As the field runs to a larger scale and closer to the confining scale, it receives higher-order corrections from the Kähler potential. These corrections cannot prevent the runaway in a supersymmetric theory, but they can in a theory with SUSY breaking. One can see this by noticing that the higher-order Kähler corrections contribute to the scalar potential in the form of Eq.~\eqref{eq:treeB}.

The effect of these higher-order Kähler corrections for the phase diagram at zero chemical potential case was explored in~\cite{deLima:2023ebw}. These higher-dimensional operators are expected to be weakly coupled and thus should not strongly affect the zero chemical potential phase diagram. However, because of the flat direction, these contributions become the dominant ones and can thus stabilize the vacuum in the low-energy theory, even for small coefficients. The leading dimension-six Kähler potential operators that are generated by the strong dynamics at low energies have the form of\footnote{The normalization of the Wilson coefficients with $N_f$ is chosen such that tree- and loop-level self-energy diagrams are finite in the large $N_f$ limit.}
\begin{align} \nonumber
-\Lambda^2 K_6 & =\frac{c_{M_1}}{N_f^2} \left[\operatorname{Tr}\left(M^{\dagger} M\right)\right]^2+\frac{c_{M_2}}{N_f} \operatorname{Tr}\left(M^{\dagger} M M^{\dagger} M\right)+\frac{c_B}{N_f}\left[\left(B^{\dagger} B\right)^2+\left(\widetilde{B}^{\dagger} \widetilde{B}\right)^2\right] \\ \nonumber
& +\frac{c_{B \widetilde{B}}}{N_f}\left(B^{\dagger} B\right)\left(\widetilde{B}^{\dagger} \widetilde{B}\right)+\frac{c_{M B}}{N_f^2} \operatorname{Tr}\left(M^{\dagger} M\right)\left(B^{\dagger} B+\widetilde{B}^{\dagger} \widetilde{B}\right) \\
& +\frac{c_{B M M B}}{N_f}\left(B^\dagger M^\dagger M B+\widetilde{B} M M^{\dagger} \widetilde{B}^{\dagger}\right) \, .
\label{eq:Kähler}
\end{align}
These operators all respect the parity invariance of the UV theory.

The values of these couplings are fixed in the actual theory (just like those of the low-energy superpotential), but we here treat them as free parameters, given that nonperturbative matching is needed to know their values. The leading order scalar potential that arises from these operators is given in App.~\ref{ap:eqs}. In addition to the contribution to the scalar potential, these operators also modify the kinetic terms of the (s)baryons and (s)mesons, but those effects are not important for this analysis, as we discuss in App.~\ref{appendix1}. The only contribution from the non-trivial kinetic terms comes from positivity conditions on the two-derivative operator, which we enforce for all configurations explored in this work.

The effect of these higher-dimensional operators can be illustrated by considering the direction $M=\widetilde{B}=0$, where only one operator contributes to the scalar potential:
\begin{align}
\label{eq:bpotential}
    V_{M=0,\widetilde{B}=0} =  m_{B , \widetilde{B}}^2\,b^{2} + \frac{1}{(N_{c}+1)\Lambda^{2}}c_{B}\,m_{3/2}^{2}\,b^{4} \, .
\end{align}
From Eq.~\eqref{eq:bpotential} we can see that the runaway direction is lifted as long as $c_{B}$ is positive. This highlights that the Kähler coefficients have a preferred sign at leading order. This occurs because of the truncation in $1/\Lambda^2$. If the $c_{i}$ coefficients are negative, there will be runaway directions in the scalar potential at this fixed order. This is not intrinsically problematic, but it means that perturbation theory in $1/\Lambda$ does not describe the theory. Namely, any small fluctuation away from the deep IR would drive the theory to $\Lambda$. We here assume that the deep IR is perturbative, which implies that the Wilson coefficients in Eq.~\eqref{eq:Kähler} are positive.  Under the assumption that all Kähler coefficients are positive, we can study the phase diagram of this theory. 

For the remainder of this work, we restrict our attention to the case $N_{c}=3$ and $N_f = 4$ and perform a numerical study of the vacuum structure in the presence of finite $\mu_B$. The higher-order Kähler terms—assumed positive—enable us to stabilize the potential and explore the full phase diagram under perturbative control.

\section{Phases of four-flavor ASQCD}
\label{sec:cons}

Because of the complexity of the additional Kähler operators needed to stabilize the runaway direction and the tachyonic mass for the sbaryons, the possible breaking patterns have many possible directions described by Eq.~\eqref{eq:gen}. Most of the breaking patterns cannot be described analytically, and it is necessary to use numerical analysis. In our numerical analysis, we initially allowed for the vacuum to take the most general form. However, the only symmetry-breaking patterns that the numerical analysis found are described by the simpler form:
\begin{align}
B=\left(\begin{array}{c} \label{eq:ansatzagain}
b \\
0 \\
0 \\
0
\end{array}\right) \, , \quad  \,  \widetilde{B}=\left(\begin{array}{c}
\bar{b} \\
0 \\
0 \\
0
\end{array}\right) \, , \quad \,  M=\left(\begin{array}{llll}
x & & & \\
& v & & \\
& & v & \\
& & & v
\end{array}\right) \, ,
\end{align}
with only 4 nontrivial directions $(x, v, b, \bar{b})$. This helps us to describe some analytical properties of the phase diagram since all transitions are accounted for by this reduced 4-dimensional field space. Focusing only on the $(x, v, b, \bar{b})$ directions, the potential still has a complicated form:
\begin{align}
 V_{\rm ASQCD}  &= V_{0} + V_{K} \, , \\ \nonumber
 V_{0} &= \frac{9 \,\kappa ^4 }{256 \pi ^4} \,m_{3/2}^2\, \left(x^2 + 3 \, v^2\right)+\left(\frac{9\,\kappa^4}{64 \pi^4} \,m_{3/2}^2 - \, \mu_{B}^2\right) \,\left(b^2 +\bar{b}^2 \right) + \kappa^2 \,x^2 \,\left(b^2+\bar{b}^2\right)  \\ 
 &- \frac{9\, \kappa ^3 }{16\pi^2}\,m_{3/2}\, x\, b\, \bar{b} - 2\,\kappa \, \lambda \,b \,\bar{b} \, v^3 +2 \,\lambda\,  m_{3/2}\, v^3\, x+\lambda ^2\, v^6+3 \,\lambda^2 \,v^4\, x^2 + \kappa^2\, b^2\,\bar{b}^2 \, , 
\end{align}
where we have chosen units with $\Lambda=1$. In our analysis, we include only the leading order contribution from the Kähler operators in Eq.~\eqref{eq:Kähler}, and $V_{K}$ is the leading scalar potential from the Kähler corrections with the explicit dependence given in App.~\ref{ap:eqs}. 

We investigate the scalar potential both numerically and analytically following a similar approach as~\cite{deLima:2023ebw}. Identifying the possible local minima numerically, we have found several novel symmetry-breaking patterns that depend on the parameters of the low-energy theory. The novel phases that we have identified are listed in Table~\ref{table:vacs}. Two non-trivial phases can be described analytically, and their approximate scalings are given in the last column of Table~\ref{table:vacs}. 
Two of the vacua identified spontaneously break $U(1)_B$ but have a residual $U(1)_\text{res}$. In both cases, the residual $U(1)$ is a linear combination of baryon number and of the generator
\begin{equation}
    T_\text{res} \propto \text{diag}\left(-3,1,1,1 \right)
\end{equation}
from one of the non-Abelian groups. 
For {\color{DarkerPurple}$SU(3)_V\times U(1)_\text{res}$},  $T_\text{res}$ is from the diagonal combination of $SU(4)_L$ and $SU(4)_R$, while for {\color{DarkerBlue}$SU(3)_{L/R}\times SU(4)_{R/L} \times U(1)_{\rm res}$}, $T_\text{res}$ is from the factor of $SU(4)$ that is broken down to $SU(3)$. We further note that the vacuum {\color{DarkerBlue}$SU(3)_{L/R}\times SU(4)_{R/L} \times U(1)_{\rm res}$} spontaneously breaks P, while the last vacuum {\color{DarkerOrange}$SU(3)_L \times SU(3)_R$} may also spontaneously break P, if $b = \bar{b}$. For both cases, one has two degenerate vacua related to each other by P transformation $B \leftrightarrow \widetilde{B}^*$ and $\psi_{B} \leftrightarrow \psi^c_{\widetilde{B}}$.

The solutions with analytical understanding show an interesting scaling that was pointed out in~\cite{deLima:2023ebw} at zero chemical potential. The VEVs are proportional to the inverse of the couplings $\lambda$ or $c_B$. This means that if we are in a naively very perturbative regime with $\lambda, c_B \ll 1$, then the VEVs are larger than $\Lambda$ and thus outside the regime of validity of the low energy effective field theory. Therefore, these vacua are only under control for moderate values of $c_B$ and $\lambda$. This qualitative behavior is also present in the vacua that can only be found numerically.
\begin{table}[t!]
\begin{tabular}{|c|c|c|c|}
\hline
Low energy symmetry                                              &  \, Configuration \,  & \, Goldstone bosons \, & \, Scaling of the VEV  \,                                \\ \hline
{\color{DarkerRed}$SU(4)_{L} \times SU(4)_{R} \times U(1)_{B}$}          & $(0, 0, 0, 0)$                                                                 & 0                & $-$                                               \\ \hline
{\color{DarkerGreen}$SU(4)_{V} \times U(1)_{B}$ }                          & $(v, v, 0, 0)$                                                                 & 15               & $\frac{v^2}{\Lambda^2} \sim \frac{m_{3/2}}{\lambda}$ \\ \hline
{\color{DarkerPurple}$SU(3)_V\times U(1)_{\rm res}$}                            & $(x, v, b, b)$                                                                 & 22               & $\sim$                                                  \\ \hline
{\color{DarkerBlue}$SU(3)_{L/R}\times SU(4)_{R/L} \times U(1)_{\rm res}$} & \begin{tabular}[c]{@{}c@{}}$(0, 0, b, 0)$\\  $(0, 0, 0, \bar{b})$\end{tabular} & 7                & $\frac{b^2}{\Lambda^2} \sim \frac{1}{c_B}$           \\ \hline
{\color{DarkerOrange}$SU(3)_L \times SU(3)_R$   }                           & \begin{tabular}[c]{@{}c@{}}$(x, 0, b, \bar{b})$\\ $(x, 0, b, b)$\end{tabular}  & 14               & $\sim$                                                  \\ \hline
\end{tabular}
\caption{\label{table:vacs}Description of the different vacuum configurations in terms of $(x,v,b,\bar{b})$ defined in Eq.~\eqref{eq:ansatzagain}. The second line is the true vacuum with zero chemical potential to leading order in $m_{3/2}/\Lambda$~\cite{Murayama:2021xfj}. Including NLO Kähler corrections, the vacua in the first and third rows arise~\cite{deLima:2023ebw}.  The last two rows are only present with a nonzero chemical potential. The  {\color{DarkerBlue}$SU(3)_{L/R}\times SU(4)_{R/L} \times U(1)_{\rm res}$} vacuum spontaneously breaks parity. The  {\color{DarkerOrange}$SU(3)_L \times SU(3)_R$} vacuum can be either $P$-preserving or $P$-breaking depending on the low-energy couplings. We also highlight the scaling with some of the low energy couplings for vacua that have analytic solutions. }
\end{table}

The non-trivial scaling of the VEVs with low-energy couplings creates an interesting phenomenon when considering specific parameter points. Consider the following scenario at zero chemical potential: the theory in the deep IR is studied at a given value of $\lambda=\lambda_0$, which is small, but not small enough such that the VEV is greater than $\Lambda$. The global\footnote{Throughout this work, a global minimum refers to the lowest vacuum energy within the effective (s)hadron theory.} minimum of the potential is a QCD-like vacuum. We next turn on the Kähler terms in Eq.~\eqref{eq:Kähler} with small coefficients $c_W$. This can induce a new minimum at field values $\gg \Lambda$. The QCD-like vacuum, however, is still a minimum and is approximately unchanged. The new vacuum makes the interpretation of the potential minimization difficult. In principle, VEVs above $\Lambda$ are outside the regime of validity of the EFT and cannot be trusted. As long as there is no path through field space within the EFT where the field runs towards that non-perturbative vacuum, the calculation of the QCD-like minimum should still be valid. However, this analysis can be more nuanced because as $c_W$ increases, the new vacuum gets closer to $\Lambda$, and it becomes difficult to know when to begin to trust the existence of that vacuum as the Wilson coefficients are varied.

\begin{figure}[thb!]
\centering
 \resizebox{0.49\linewidth}{!}{ \includegraphics{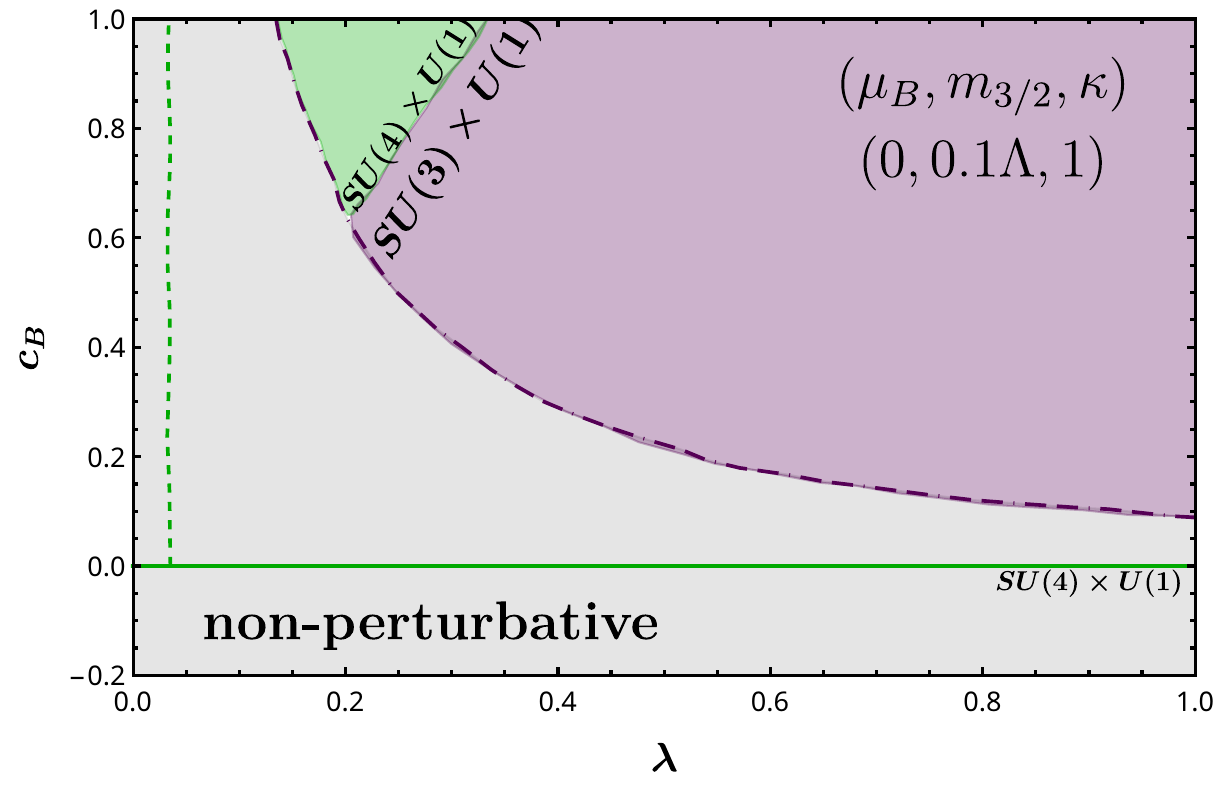}} 
  \resizebox{0.49\linewidth}{!}{ \includegraphics{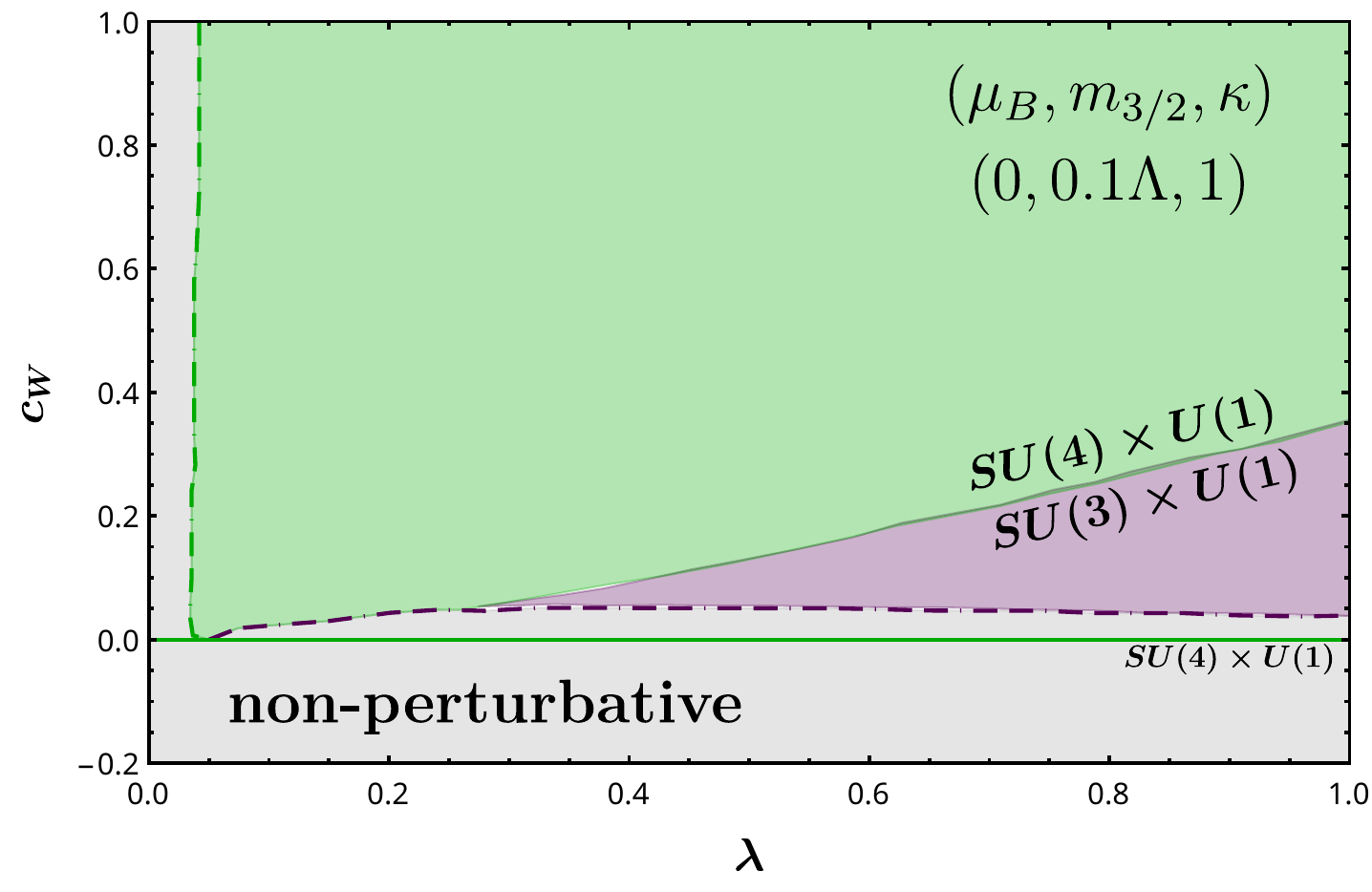}} 
\caption{\label{fig:zero} Phase diagram with Kähler corrections at zero chemical potential for a specific parameter point: $m_{3/2}=0.1\,\Lambda$, and $\kappa=1$. On the left,  only $c_B$ from Eq.~\eqref{eq:Kähler} is nonzero, while on the right, we include all Wilson coefficients with the same value, $c_i\equiv c_W$. Notice that in the limit with no higher order Kähler terms, the vacuum is QCD-like {\color{DarkerGreen}$SU(4)_{V} \times U(1)_{B}$}. As the Kähler coefficient grows, the {\color{DarkerPurple}$SU(3)_V\times U(1)_{\rm res}$} vacuum appears at large field values, and eventually becomes perturbative. Any parameter point where a local minimum appears for field values bigger than $\Lambda$ is classified as non-perturbative. The other colored regions correspond to the global minimum potential when no vacua have VEV $>\Lambda$, which for this choice of parameters are either {\color{DarkerGreen}$SU(4)_{V} \times U(1)_{B}$ }  or {\color{DarkerPurple}$SU(3)_V\times U(1)_{\rm res}$}. It is worth noting that the region of small $\lambda$ is also out of theoretical control because the VEV in the QCD-like vacuum becomes $>\Lambda$. 
}
\end{figure}

When we have a vacuum with VEV greater than $\Lambda$, or if a runaway direction exists in the potential, we cannot describe the fate of the theory with certainty. In this work, we take the conservative approach where if for a particular choice of coupling constants the potential minimization calculation reveals a vacuum with VEV $> \Lambda$, then we classify the corresponding parameter point as ``non-perturbative.'' To illustrate this, we first consider $\mu_B=0$, fixing $m_{3/2}$ and $\kappa$, and examining two possibilities for the Kähler operators from Eq.~\eqref{eq:Kähler}. 
In the first scenario, only 
$c_B$ is nonzero, allowing us to isolate its effects. In the second, all dimension-six operators are turned on with the same value, $c_W$.  The phase diagram as a function of the Kähler coefficients and the superpotential coupling $\lambda$ is shown in Fig.~\ref{fig:zero}. We see that a significant portion of the parameter space is in our conservative ``non-perturbative'' classification. Finally we note that the vacuum energy and VEV of the QCD-like {\color{DarkerGreen}$SU(4)_{V} \times U(1)_{B}$} vacuum is approximately insensitive to the Kähler operators, and the reason the left and right panels of Fig.~\ref{fig:zero} look so different is because the exotic {\color{DarkerPurple}$SU(3)_V\times U(1)_{\rm res}$} vacuum is significantly sensitive to those operators.

\section{Phase diagram for four-flavor ASQCD}
\label{sec:phase}

We now have all the elements to explore the finite chemical potential behavior of four-flavor ASQCD. At low energies, the theory is weakly coupled, and the phase of the theory is dominated by the scalar dynamics. We can then explore the dynamics for finite baryon chemical potential below the confinement scale $\mu_B < \Lambda$, and construct its phase diagram.

To explore the phase of the theory, we need to fix the low-energy constants. Unfortunately, unlike real QCD, we do not have experimental guidance on their values, and in turn, we treat them as free parameters and try to explore all sensible values. The free parameters include the superpotential couplings $\kappa$ and $\lambda$ [see~Eq.~\eqref{eq:superpotential}], and the six dimension-six Kähler potential terms [see Eq.~\eqref{eq:Kähler} and App.~\ref{appendix1}]. There is also, of course, $m_{3/2}$ that controls the importance of SUSY breaking effects. 

Because of the different interpretations of the parameters of the theory, it is instructive to explore the $\mu_{B}$ vs.~$m_{3/2}$ plane with the other couplings marginalized. This plane highlights the impact of SUSY breaking on the phase of the theory. As a guide, we marginalize the free parameters using naive dimensional analysis (NDA)~\cite{Weinberg:1978kz,Luty:1997fk,Cohen:1997rt} to estimate the values of the couplings at the confinement scale $\Lambda$. We can also use the leading order renormalization group (RG) running to evolve from $\Lambda$ down to a lower scale $\mu_{\rm RG}$. The couplings are given by 
\begin{align}
\kappa(\mu_{\rm RG}) =\frac{4\pi}{1-18\log(\frac{\mu_{\rm RG}}{\Lambda})} \, ,\quad \, \lambda(\mu_{\rm RG}) = (4\pi)^{2} \left(\frac{\mu_{\rm RG}}{\Lambda} \right) \, ,\quad \,   c_{B}(\mu_{\rm RG}) = (4\pi)^{3} \left(\frac{\mu_{\rm RG}}{\Lambda} \right)^{2} \, ,
\label{eq:rg}
\end{align}
with similar expressions for the coefficients of the other dimension-six Kähler operators. 

We first explore the dynamics when we strictly follow NDA expectations, taking $\mu_{\rm RG}$ to be the larger of $\mu_B$ and $m_{3/2}$ at a given parameter point. We restrict our analysis so that both scales are smaller than the confining scale $\Lambda$. We also require that the theory remains in the perturbative regime in terms of the low energy EFT expansion. This restricts the possible values of the dimensionful parameters to be smaller than $\mu \approx (4\pi)^{-1}\Lambda$. As in Fig.~\ref{fig:zero}, we consider two possibilities for the Kähler operators from Eq.~\eqref{eq:Kähler}. The first is $c_B$, taking its NDA value and the rest zero, and the second is with all coefficients set to their NDA value.  

\begin{figure}[t!]
 \resizebox{0.49\linewidth}{!}{ \includegraphics{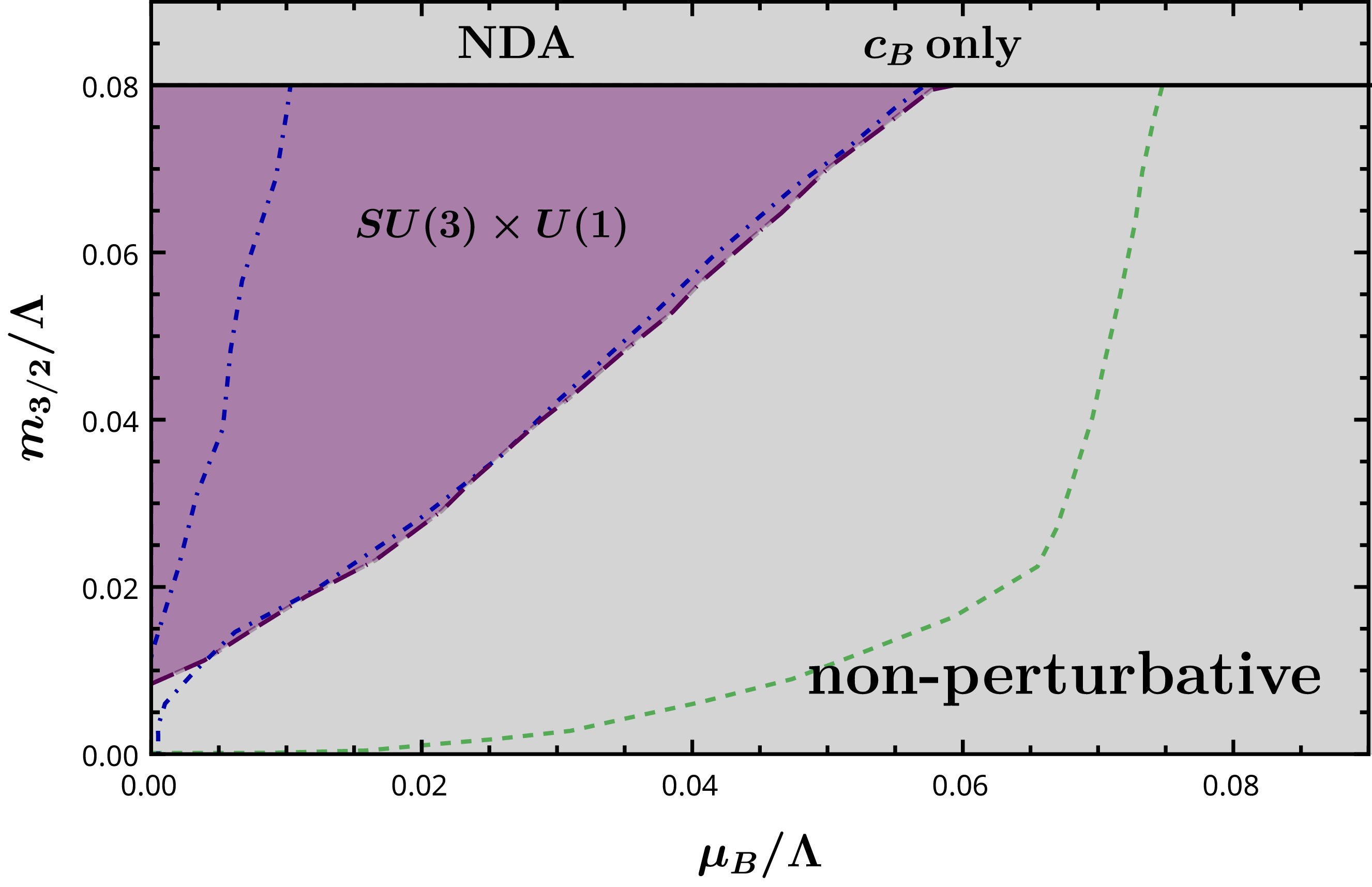}} 
  \resizebox{0.49\linewidth}{!}{ \includegraphics{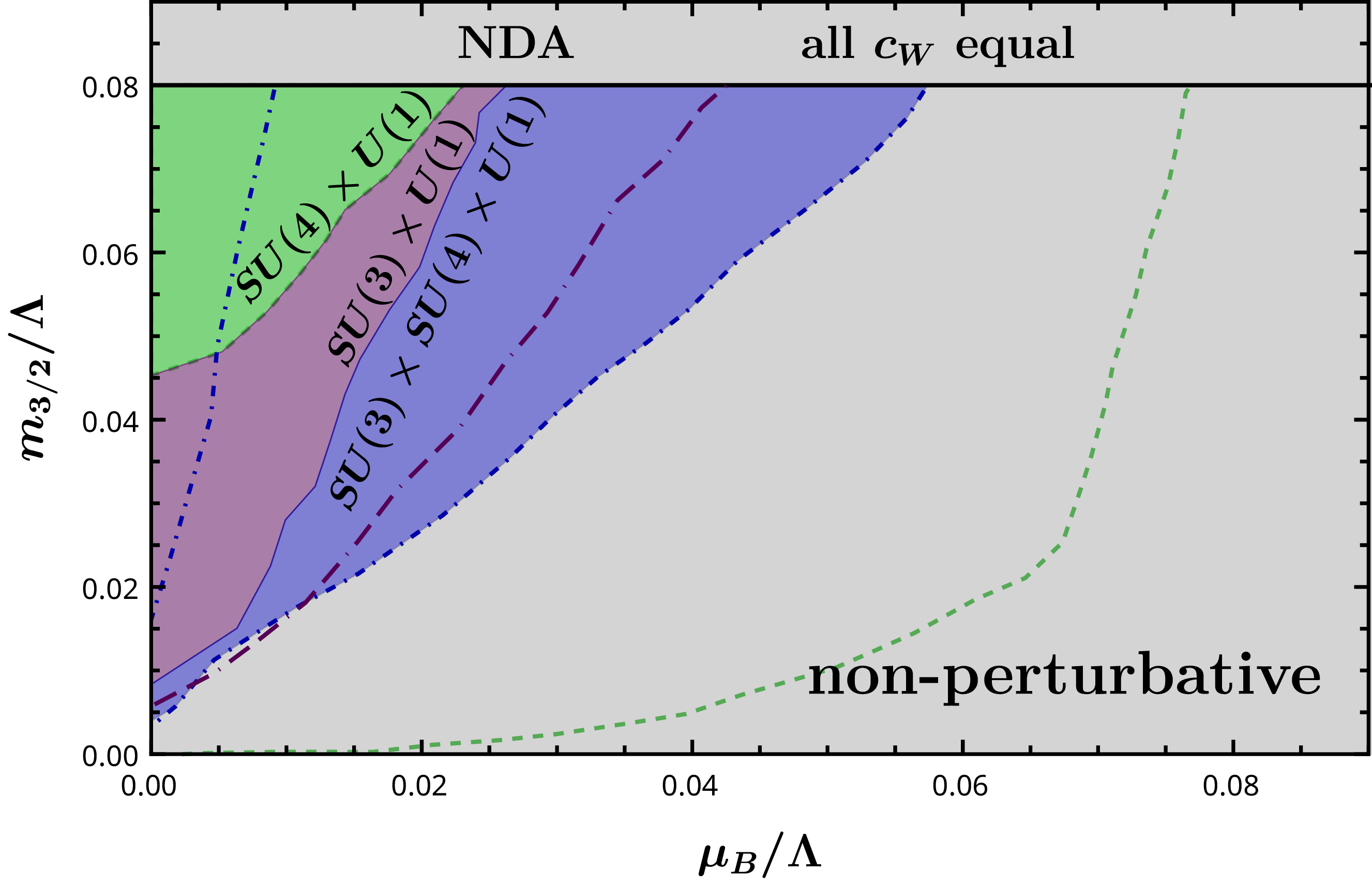}} 
\caption{Phase diagram in the plane of SUSY breaking $m_{3/2}$ vs.~baryon chemical potential $\mu_B$ for couplings following NDA expectations. On the left, we include only one Kähler operator, while on the right, we include all dimension-six operators with the same coefficient. As in Fig.~\ref{fig:zero}, the ``non-perturbative'' region means that there is some VEV $>\Lambda$, and the colored regions are the global minimum. The lines represent boundaries when a particular phase ceases to exist. 
In this figure and throughout this article, we classify the vacuua from Table~\ref{table:vacs} as follows:  the {\color{DarkerGreen}$SU(4)_{V} \times U(1)_{B}$ } vacuum is colored in green and with dashed curves. The {\color{DarkerPurple}$SU(3)_V\times U(1)_{\rm res}$} vacuum is colored with purple and with dot-dash-dash curves. The {\color{DarkerBlue}$SU(3)_{L/R}\times SU(4)_{R/L} \times U(1)_{\rm res}$} vacuum  is colored with blue and with dot-dashed curves.
}
\label{fig:NDA}
\end{figure}

With this setup, we can see the phase diagram in Fig.~\ref{fig:NDA}. The choice where only $c_{B}$ is non-zero is on the left, while the case with all equal dimension-six Wilson coefficients $c_{W}$ is shown on the right. In this figure and throughout this article, we use the convention where the colored regions are global minima, and the lines represent the existence or destabilization of local minima. To differentiate the types of phases, we dash each minimum differently and with different colors, with the following convention:
\begin{itemize}
    \item {\color{DarkerGreen}$SU(4)_{V} \times U(1)_{B}$ } is colored in green and with dash curves,
    \item {\color{DarkerPurple}$SU(3)_V\times U(1)_{\rm res}$   } is colored with purple and with dot-dash-dash curves,
    \item {\color{DarkerBlue}$SU(3)_{L/R}\times SU(4)_{R/L} \times U(1)_{\rm res}$} is colored with blue and with dot-dashed curves,
    \item {\color{DarkerOrange}$SU(3)_L \times SU(3)_R$} is colored in orange and with dot-dot-dash curves.
\end{itemize}

From Fig.~\ref{fig:NDA}, we can see some interesting features. In the plot on the left, we see that the global vacuum is not the QCD-like chiral symmetry breaking one, even for small $\mu_B$. This is not the expectation when $c_B=0$. Once the other Kähler operators start to contribute as shown in the plot on the right, the QCD-like vacuum comes back to be the global minimum for small $\mu_B$ and relatively large $m_{3/2}$. This feature demonstrates the sensitivity of the phase of the theory to low energy constants. While our objective here is to map and classify all possible phases, we do not focus on one specific set of low-energy constants. 

As discussed in Section~\ref{sec:cons}, the region labeled ``non-perturbative'' is defined when there exists a vacuum configuration with a field VEV larger than the confinement scale $\Lambda$, or when there is a runaway direction that is not stabilized. For much of the regions labeled non-perturbative in Fig.~\ref{fig:NDA}, there \textit{are} locally stable minima that are perturbative. 

The QCD-like chiral symmetry breaking VEV scales like $1/\sqrt\lambda$, while the  {\color{DarkerPurple}$SU(3)_V\times U(1)_{\rm res}$} VEV scales inversely with the Wilson coefficients.\footnote{This scaling is derived from the behavior of the numerical solutions found in the scans.} This means that these solutions are starting at large field values and being moved inwards as the coupling increases. Since the hierarchy between them is fixed by both NDA and perturbativity, the {\color{DarkerPurple}$SU(3)_V\times U(1)_{\rm res}$} vacuum is more likely to be deeper as $c_B \ll \lambda$ in this region. The same is not necessarily true once other Kähler operators are included. In this case, one observes the overall trend that the majority of the non-trivial part of the phase diagram is dominated by the {\color{DarkerBlue}$SU(3)_{L/R}\times SU(4)_{R/L} \times U(1)_{\rm res}$} with the spontaneous breaking of parity.

\begin{figure}[t!]
\resizebox{0.3285\linewidth}{!}{ \includegraphics{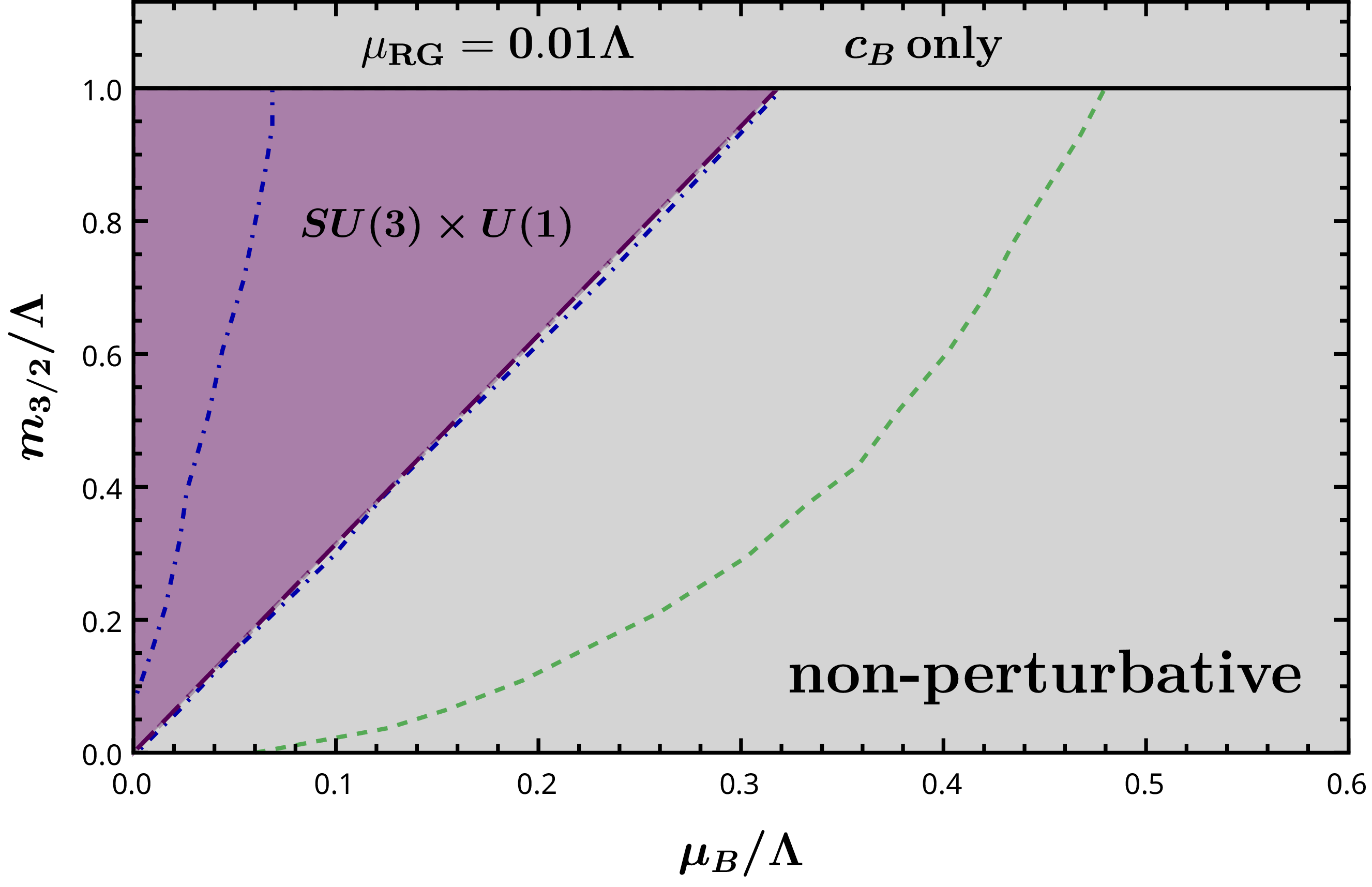}} 
\resizebox{0.3285\linewidth}{!}{ \includegraphics{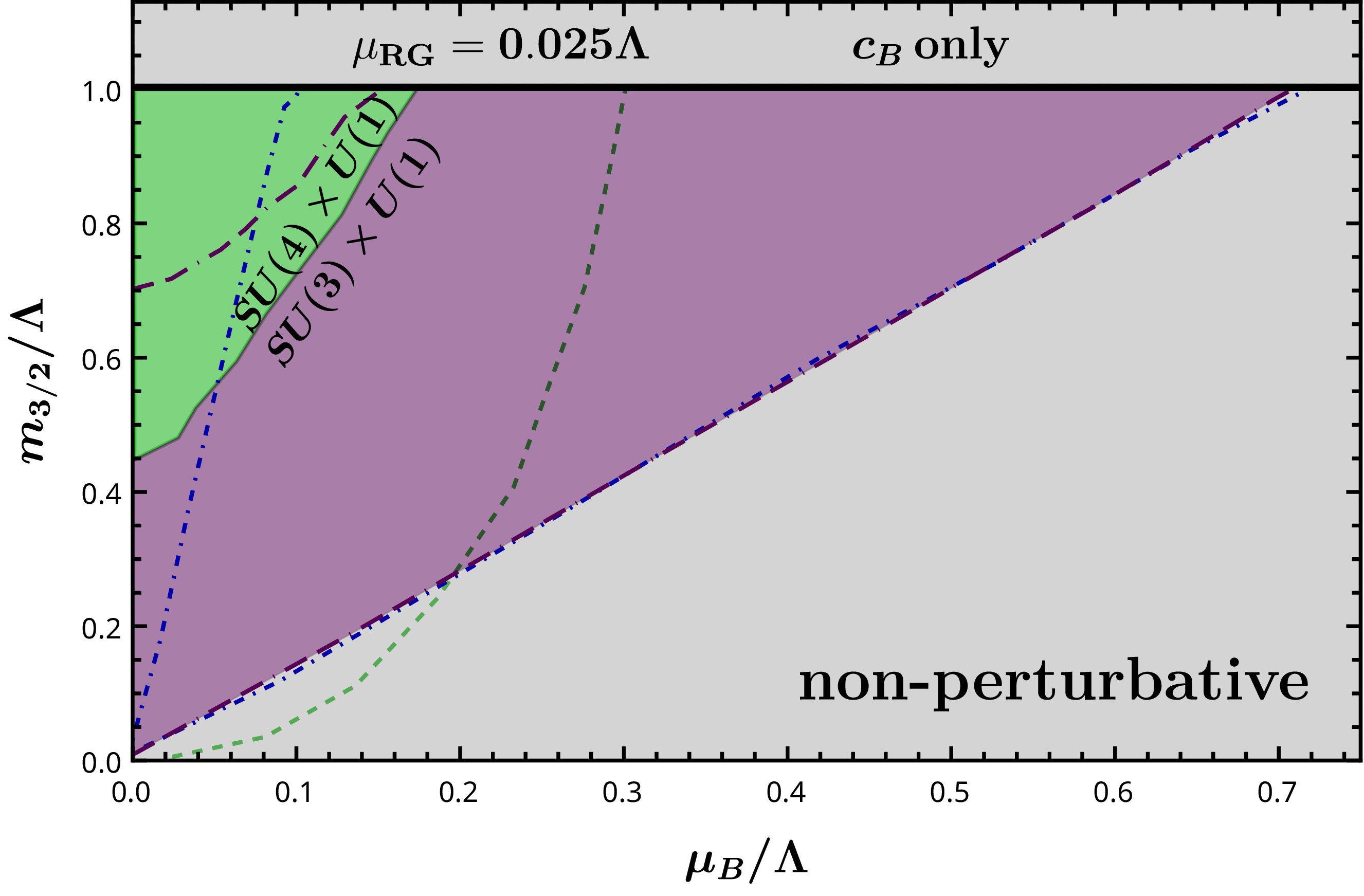}} 
\resizebox{0.3285\linewidth}{!}{ \includegraphics{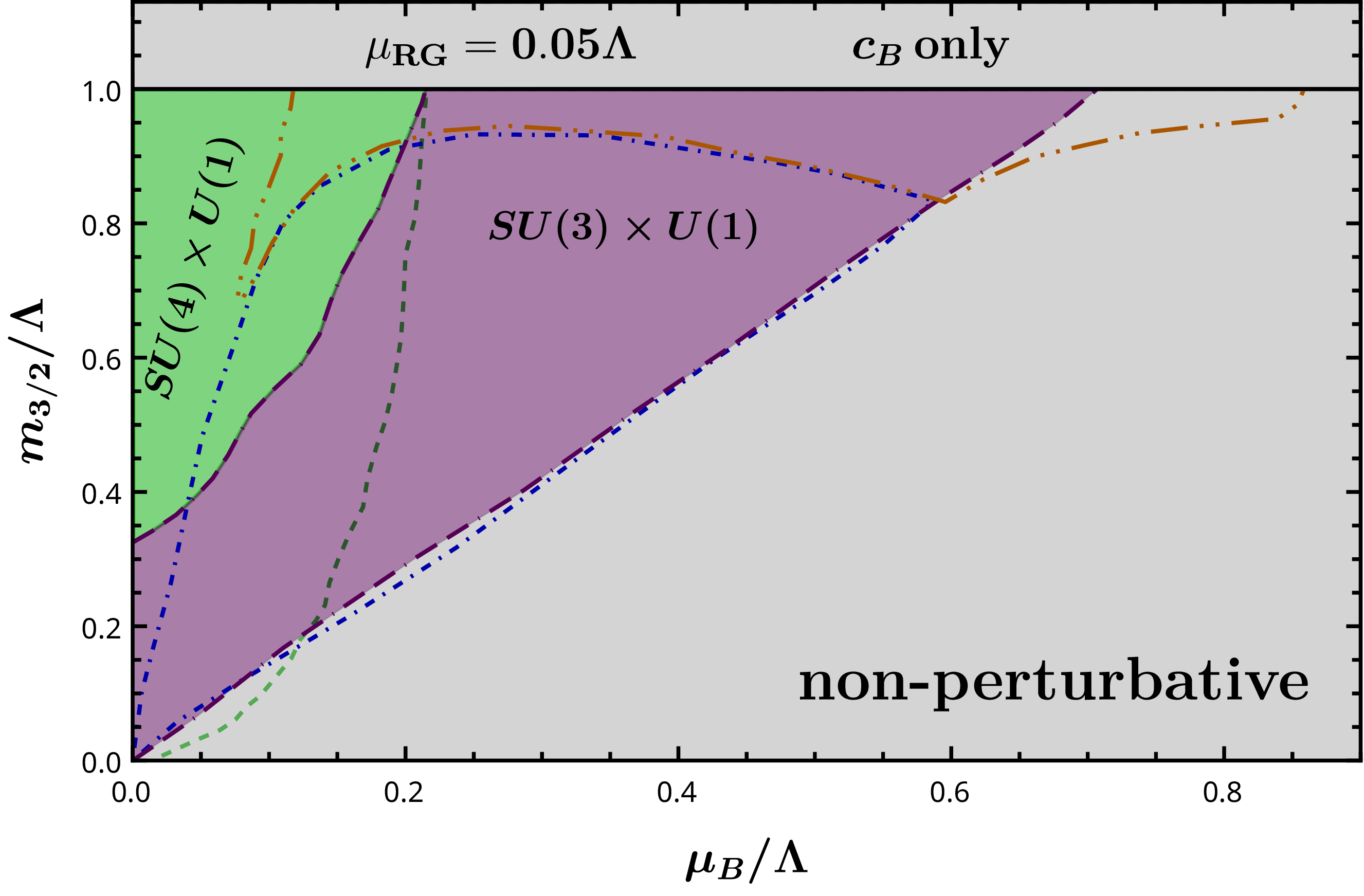}} 
\resizebox{0.3285\linewidth}{!}{ \includegraphics{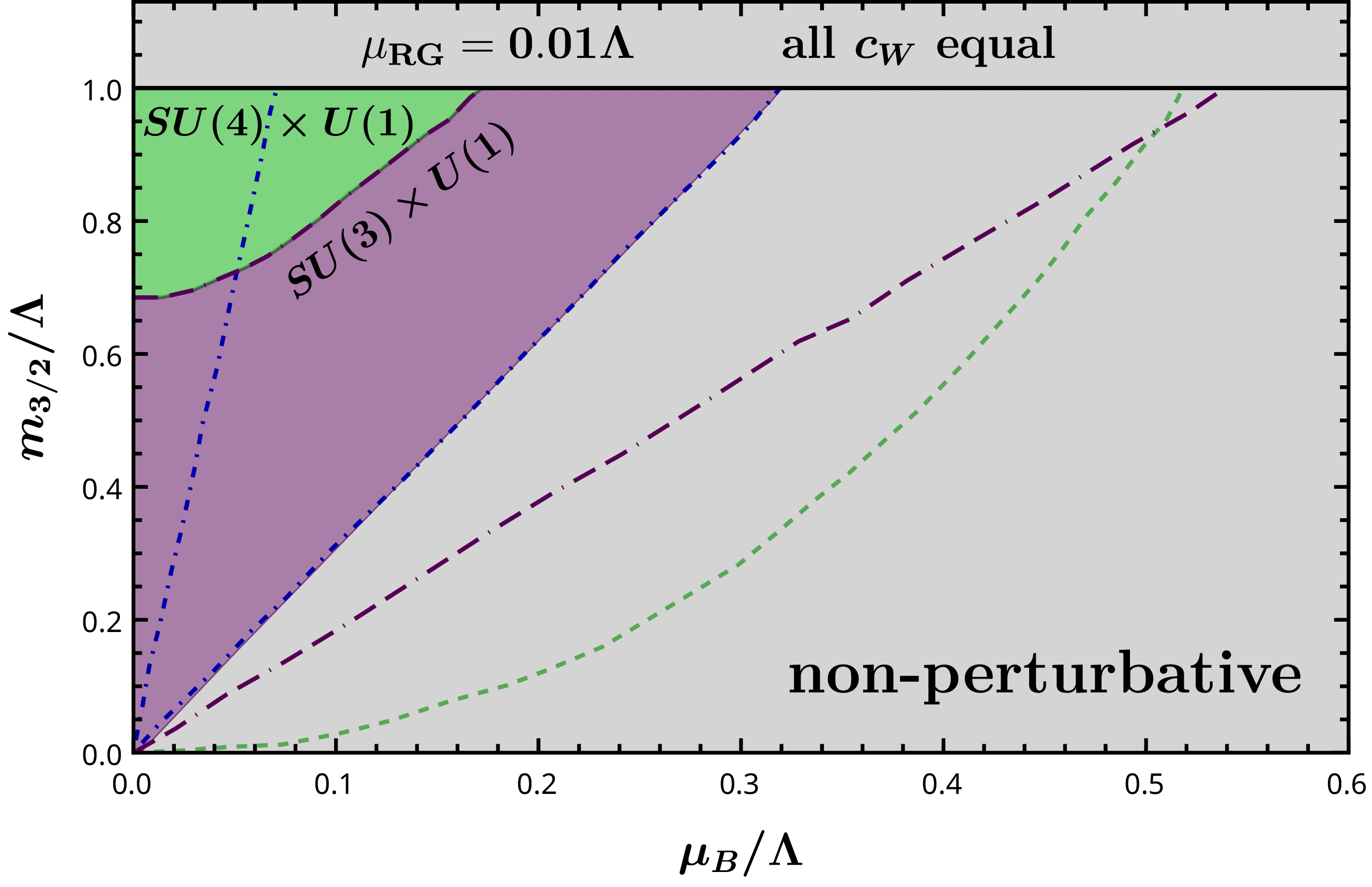}} \resizebox{0.3285\linewidth}{!}{ \includegraphics{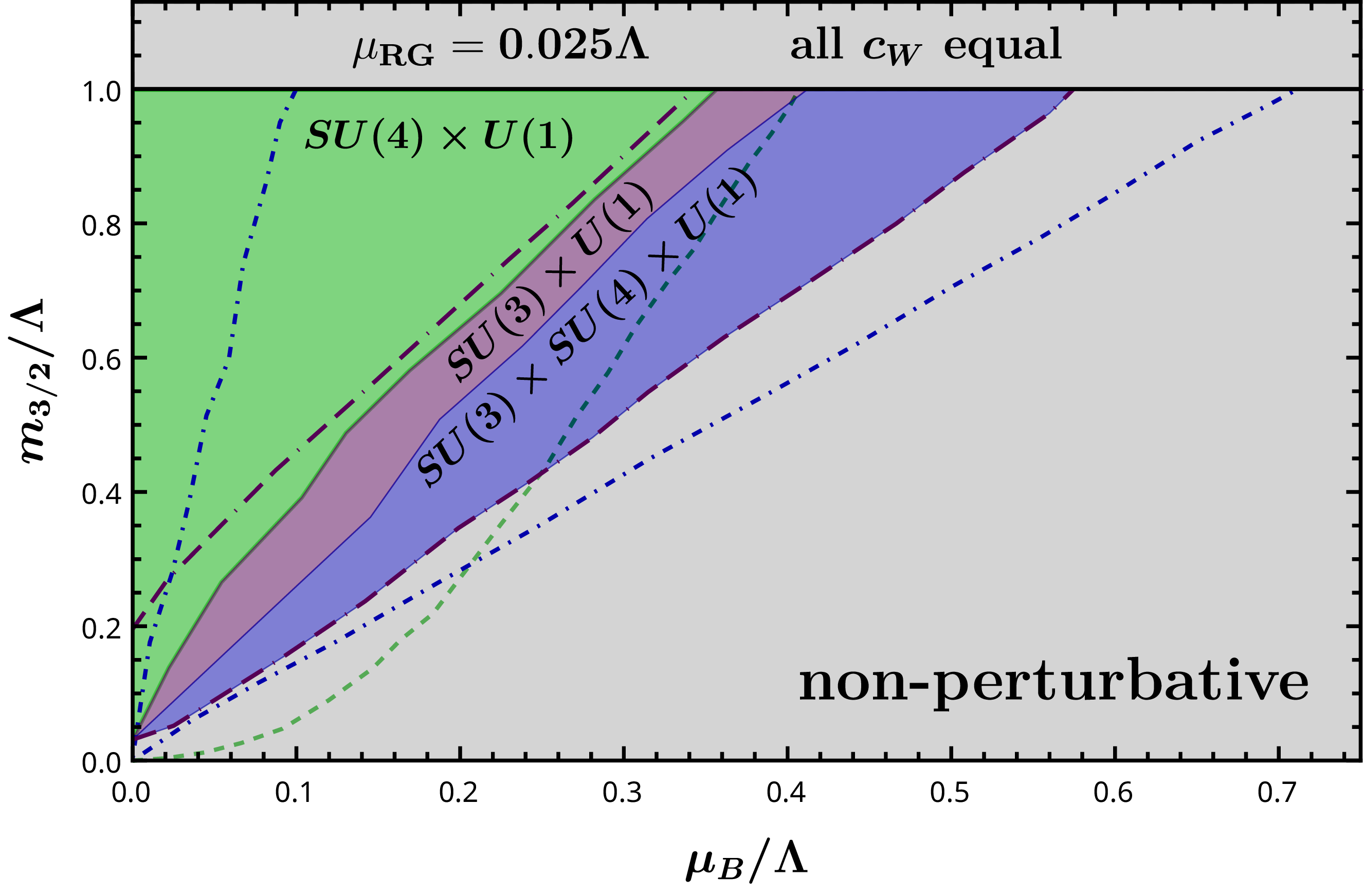}} 
 \resizebox{0.3285\linewidth}{!}{ \includegraphics{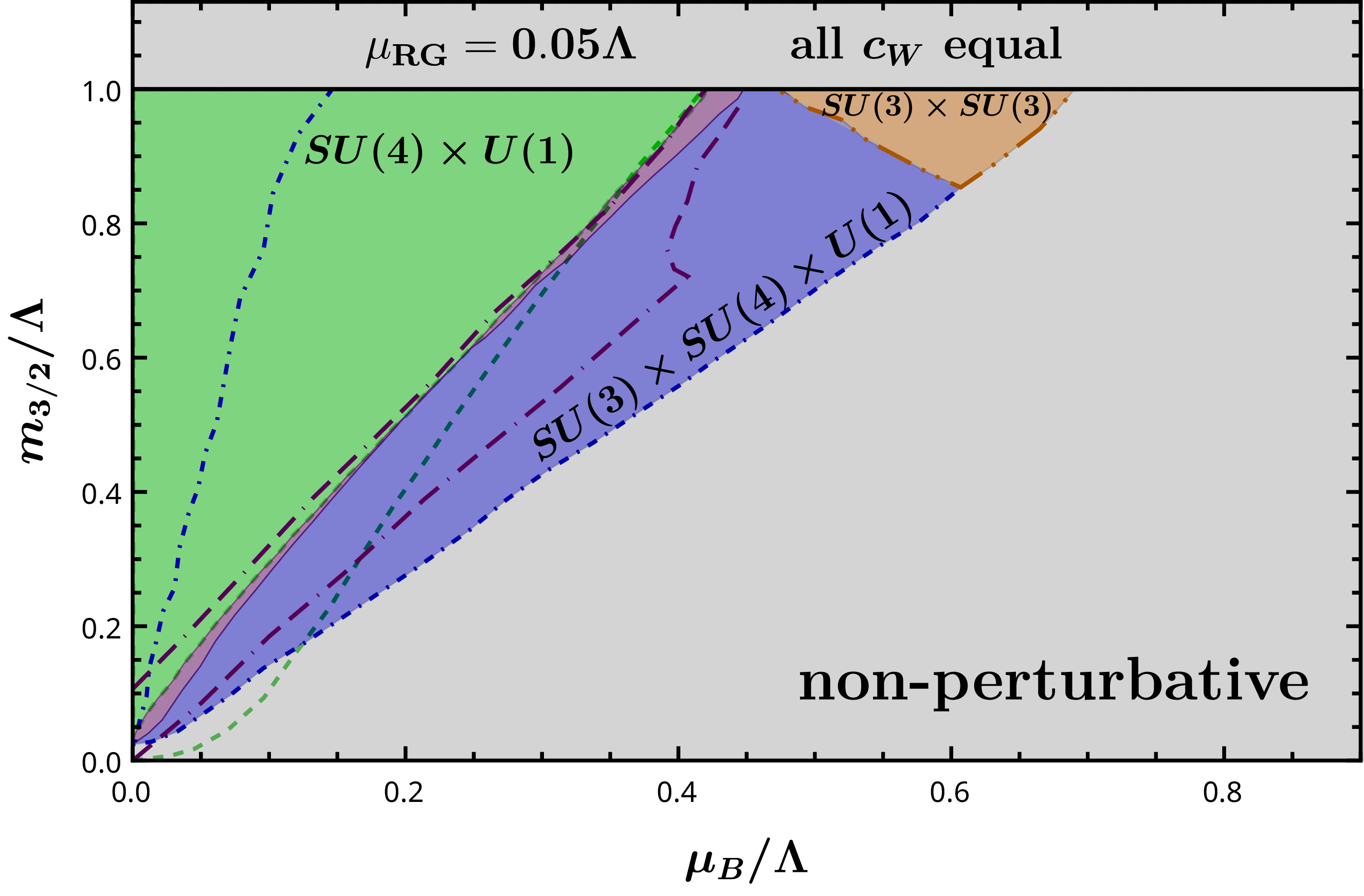}} 
\caption{  Phase diagram in the plane of SUSY breaking ($m_{3/2}$) vs.~baryon chemical potential ($\mu_B$) for the couplings fixed at one specific value of $\mu_{\rm RG}$ [see Eq.~\eqref{eq:rg}]. On the top row, we consider only $c_{B}$ nonzero, while on the bottom row, we include all dimension-six operators with the same coefficient. The {\color{DarkerOrange}$SU(3)_L \times SU(3)_R$} vacuum is colored in orange and with dot-dot-dash curves. The other configurations follow the same style as in the other figures. }
\label{fig:FIX}
\end{figure}

We can also explore a different prescription for marginalizing the low-energy constants. We preserve NDA scaling but choose a fixed RG scale such that the couplings are the same for all values of $m_{3/2}$ and $\mu_B$ in the plane. This allows us to extend the plot axes up to $\Lambda$, and the phase diagram for different values of couplings can be seen in Fig.~\ref{fig:FIX}. In the top row, we turn on only the $c_B$ operator, while in the bottom row, all the Kähler operators in Eq.~\eqref{eq:Kähler} are on with the same coefficient. Each column uses a different size for the Kähler operator, $c_{B/W} \simeq 0.19, 1.24, 4.96$ from left to right. 

With this different parameterization of the low energy couplings, we see a new phase with  {\color{DarkerOrange}$SU(3)_L \times SU(3)_R$} remnant symmetry. This configuration is interesting as it can either conserve or break parity depending on the couplings. There are also values of $m_{3/2}$ where there are three different perturbative global minima as we increase $\mu_B$. The significant difference between including only $c_B$ and including all the Kähler operators signals that the phase structure at finite chemical potential is quite sensitive to the strong dynamics. This is to be expected, as the only reason for the existence of a finite density equilibrium configuration is due to the effects of the higher order terms. 
This dependence on the Kähler coefficients makes it very difficult to extrapolate to large values of $m_{3/2}$, even if we assume preservation of the universality class as we flow to the non-perturbative domain.

We also explore deviations from NDA scaling of the Wilson coefficients in App.~\ref{ap:notNDA}, where we find the same phases, but without a clear preference for one over the others. The major trends are that once all Kähler operators are active, there is a stronger preference for the QCD-like vacuum at small $\mu_B$. As $\mu_B$ increases, there is in most cases\footnote{In some cases, where only $c_B$ is active, the vacuum at zero baryon chemical potential is not QCD-like and no transition occurs as $\mu_B$ increases. If a QCD-like vacuum exists, then it always transitions out of it as $\mu_B$ grows.} a change of the global-minimum phase before reaching the non-perturbative regime where our calculation breaks down. We therefore expect that this QCD-like theory with four massless flavors will undergo a phase transition in the region $0 < \mu_B < \Lambda$.

\begin{figure}[t!]
 \resizebox{0.75\linewidth}{!}{ \includegraphics{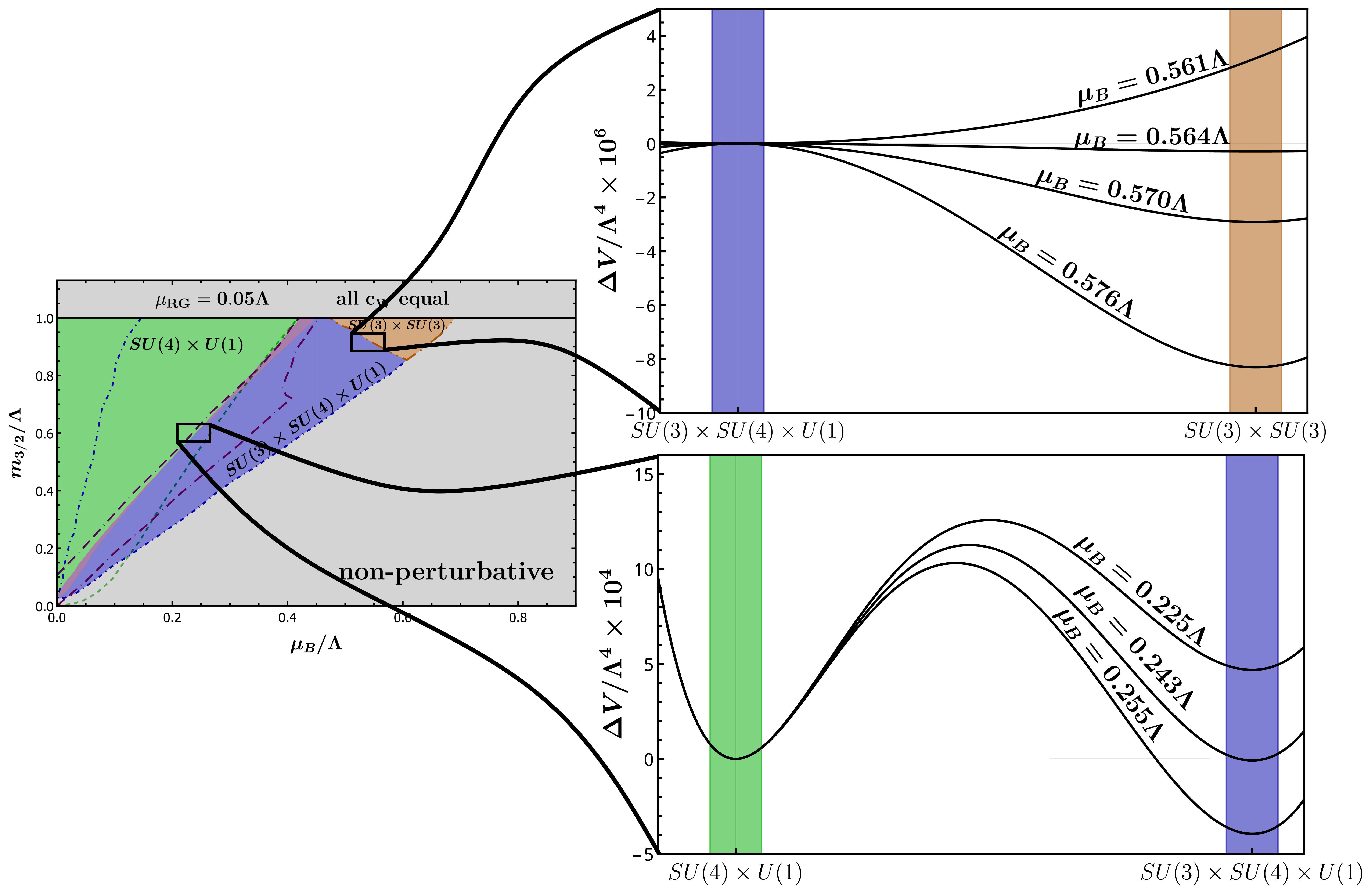}} 
\caption{Highlight of the types of phase transitions that can occur. If there is no line at the boundary of two minima in the left figure or in the previous figures, the transition is first order, while if the transition occurs on top of a line, the transition is second order. To show this, we plot the scalar potential in a straight line connecting the two vacua for the two types of transitions to demonstrate a second (first) order transition on the top (bottom). The color and dash styles are the same as in the other figures.}
\label{fig:phaseT}
\end{figure}

\section{Phase transitions}
\label{sec:transitions}

In this section, we explore the characteristics of the phase transition across the phase diagram. The only physical parameter that is externally tunable is $\mu_B$, so we explore horizontal slices of Figs.~\ref{fig:NDA},~\ref{fig:FIX}, and~\ref{fig:FIX2} to analyze the type of phase transition that occurs. Since the local minima are shown in the plot as different types of dashed lines, it is straightforward to spot the type of transition based on the boundary type. If no dashed line is present and there is a transition between two global minima, this transition is first order. This occurs because both minima coexist on both sides of the transition line, but one becomes deeper than the other as the boundary is crossed. Alternatively, if a line exists between two global minima, this indicates that the minima are only formed at the transition boundary, indicating a second order transition.

In order to display these transitions graphically, we compute a straight-line parametrization between two different global minima and examine the potential behavior in this one-dimensional field direction. Given two vacuum configurations $\boldsymbol{\xi}_1=\left(x_1,v_1,b_1,\bar{b}_1\right)$ and $\boldsymbol{\xi}_2=\left(x_2,v_2,b_2,\bar{b}_2\right)$, we can parametrize a straight line between them using a real parameter $t$ as
\begin{align}
    \boldsymbol{d}_{12}(t) = \boldsymbol{\xi}_1+t(\boldsymbol{\xi}_1-\boldsymbol{\xi}_2) \, .
\end{align}
The vector $\boldsymbol{d}_{12}(t)$ can be used then to evaluate the potential at different values of $t$, 
$V\left(\boldsymbol{d}_{12}(t)\right)$
for $0\leq t \leq 1$. This is not necessarily the path that the fields would take in the transition, but it is a simple proxy to verify the existence of barriers between two global minima and to identify the critical $\mu_B$ that triggers the transition. We can then see an example of both a first and a second order transitions in  Fig.~\ref{fig:phaseT}.

\section{Conclusions}
\label{sec:conc}

In this work, we explored the phase structure of a QCD-like theory by studying s-confining Supersymmetric QCD  deformed by anomaly mediated supersymmetry breaking in the presence of a non-zero baryon chemical potential. This framework allows for a controlled, calculable approach to strongly coupled dynamics below the confinement scale—an energy regime notoriously difficult to access using traditional methods.

We highlight that the introduction of finite baryon chemical potential in the supersymmetric theory induces instabilities, driving the theory toward runaway directions. However, we demonstrate that these instabilities can be dynamically stabilized at low energies as the theory starts to flow from the hadronic regime to the quark-gluon regime. Crucially, the stabilization arises from higher-dimensional operators in the Kähler potential, which become significant as the sbaryon vacuum expectation value approaches the confinement scale. Interestingly, this stabilization can only occur with non-zero SUSY breaking. The supersymmetric theory with non-zero baryon chemical potential has no finite ground state.  

With the theory stabilized, we systematically explored its vacuum structure as a function of both the baryon chemical potential and the SUSY breaking scale, marginalizing the low energy coupling constants using different schemes. The Lagrangian of the theory has the symmetry group  {\color{DarkerRed}$SU(4)_{L} \times SU(4)_{R} \times U(1)_{B}$} in both the high and low energy descriptions.\footnote{The supersymmetric theory has an additional global $U(1)_R$ which is explicitly broken when supersymmetry is softly broken.} The supersymmetric theory has a moduli space with various symmetry breaking patterns. The inclusion of anomaly mediated SUSY breaking at leading order induces the formation of the QCD-like vacuum that breaks the global symmetry down to  {\color{DarkerGreen}$SU(4)_{V} \times U(1)_{B}$ }. Then, a nonzero baryon chemical potential on top of that creates a rich landscape of vacua, including
\begin{itemize}
    \item {\color{DarkerPurple}$SU(3)_V\times U(1)_{\rm res}$}\,,
    \item {\color{DarkerBlue}$SU(3)_{L/R}\times SU(4)_{R/L} \times U(1)_{\rm res}$}\,, 
    \item {\color{DarkerOrange}$SU(3)_L \times SU(3)_R$}\,. 
\end{itemize}
We also mapped out the corresponding phase diagram in the plane of SUSY breaking ($m_{3/2}/\Lambda$) vs.~chemical potential ($\mu_B/\Lambda$) and found regions exhibiting both first- and second-order phase transitions. While precise statements are difficult to make given the unknown low energy constants, some general behavior can be observed. If the vacuum at small $\mu_B$ is QCD-like, then there is always a change of the global-minimum before reaching the non-perturbative regime where our calculation breaks down. We then expect that this QCD-like theory with four massless flavors undergoes a phase transition in the region $0 < \mu_B < \Lambda$. 

Overall, our work highlights the utility of using supersymmetry with anomaly mediated supersymmetry breaking as a tool in probing strongly coupled non-supersymmetric theories and opens new avenues for understanding finite-density effects in strongly coupled quantum field theories.
The inclusion of temperature to obtain the complete phase diagram is also a direct extension of this work, and will be explored further in~\cite{tempAMSB}.

\section*{Acknowledgments}
We thank Mehdi Drissi, David McKeen, David Morrissey, Hitoshi Murayama, and Georgios Palkanoglou for their helpful conversations. YB is supported by the U.S. Department of Energy under the contract DE-SC-0017647 and DEAC02-06CH11357 at Argonne National Laboratory.
This work was supported by the Natural Sciences and Engineering Research Council of Canada (NSERC). C.H.L is also supported by TRIUMF, which receives federal funding via a contribution agreement with the National Research Council (NRC) of Canada. This work was performed in part at the Aspen Center for Physics, which is supported by a grant from the Simons Foundation (1161654, Troyer) and National Science Foundation grant PHY-2210452. This research was supported by the Munich Institute for Astro-, Particle and BioPhysics (MIAPbP), which is funded by the Deutsche Forschungsgemeinschaft (DFG, German Research Foundation) under Germany´s Excellence Strategy–EXC-2094–390783311.

\appendix

\section{Higher-order Kähler Potential}
\label{ap:eqs}
The scalar potential from the higher-order Kähler potential, at leading order in the Wilson coefficient, can be written as
\begin{align}
 V_{K} &= \frac{c_{M_1}}{16}V_{M_1} + \frac{c_{M_2}}{4}V_{M_2} + \frac{c_{B}}{4}V_B + \frac{c_{B\tilde{B}}}{4}V_{B\tilde{B}} + \frac{c_{MB}}{16}V_{MB} + \frac{c_{BMMB}}{4}V_{BMMB} \, , \\ \nonumber
 V_{M_1} &=  2 b^2 \bar{b}^2 \kappa ^2 \left(3 v^2+2 x^2\right)+4 m_{3/2} x \left(3 v^2+x^2\right) \left(4 \lambda  v^3-b \bar{b} \kappa \right) \\
 &-4 b \bar{b}
   \kappa  \lambda  v^3 \left(3 v^2+5 x^2\right)+m_{3/2}^2 \left(3 v^2+x^2\right)^2+2 \lambda ^2 v^4 \left(3 v^4+26 v^2 x^2+3 x^4\right) \, , \\ 
 V_{M_2} &= 4 x^2 \left(b^2 \bar{b}^2 \kappa ^2-2 b \bar{b} \kappa  \lambda  v^3+4 \lambda ^2 v^6\right)+4 m_{3/2}x \left[x^2 \left(\lambda  v^3-b \bar{b} \kappa \right)+3
   \lambda  v^5\right] \nonumber \\
   &+m_{3/2}^2 \left(3 v^4+x^4\right) \, , \\
   V_{B} &= m_{3/2}^2 \left(b^4+\bar{b}^4\right)-4 b \bar{b} \kappa  m_{3/2} x \left(b^2+\bar{b}^2\right)+8 b^2 \bar{b}^2 \kappa ^2 x^2 \, , \\
   V_{B\tilde{B}} &= \left[m_{3/2} b \bar{b} -\kappa  x \left(b^2+\bar{b}^2\right)\right]^2 \, , \\ \nonumber
   V_{MB} &= b^4 \bar{b}^2 \kappa ^2 +\bar{b}^2 \left[\left(3 v^2+x^2\right) \left(m_{3/2}^2+\kappa ^2 x^2\right)+8
   \lambda  m_{3/2} v^3 x+\lambda ^2 v^4 \left(v^2+3 x^2\right)\right] \nonumber \\ 
   &- 2 b^3 \bar{b} \kappa  \left(m_{3/2} x + \lambda  v^3\right) -2 b \bar{b} \kappa  \left[m_{3/2} x \left(\bar{b}^2+6 v^2+2 x^2\right)+\lambda  v^3 \left(\bar{b}^2+8 x^2\right)\right] \\
   &+ b^2 \left[\kappa ^2 \left(\bar{b}^4+x^2 \left(4 \bar{b}^2+3
   v^2\right)+x^4\right)+m_{3/2}^2 \left(3 v^2+x^2\right)+8 \lambda  m_{3/2} v^3 x+\lambda ^2 v^4 \left(v^2+3 x^2\right)\right] \, , \nonumber \\ 
   V_{BMMB} &= b^4 \bar{b}^2 \kappa ^2-2 b^3 \bar{b} \kappa  \left(m_{3/2} x+\lambda  v^3\right) -2 b \bar{b} \kappa  \left(\bar{b}^2+2 x^2\right) \left(m_{3/2} x+\lambda  v^3\right) \nonumber \\
   & +b^2 \left[\kappa ^2 \left(\bar{b}^4+4 \bar{b}^2 x^2+x^4\right)+m_{3/2}^2
   x^2+2 \lambda  m_{3/2} v^3 x+\lambda ^2 v^6\right] \nonumber \\
   &+\bar{b}^2
   \left(m_{3/2}^2 x^2+2 \lambda  m_{3/2} v^3 x+\lambda ^2 v^6+\kappa ^2 x^4\right)\, ,
\end{align}
where we have set $\Lambda=1$ as our units for the system.

\section{Derivative interactions effect on the vacuum structure}
\label{appendix1}

In this appendix, we discuss the effect of higher dimensional Kähler operators on the kinetic term. In general, the Lagrangian for both the fermions and bosons receives a correction to the kinetic term proportional to the Kähler metric. For scalar fields, these corrections take the form 
\begin{align}
\mathcal{L} = g_{ij^*}(\Phi_i) \partial_\mu \Phi^{i} \partial^\mu \Phi^{j^*} - V(\Phi_i) \, ,
\label{eq:KählerMod}
\end{align}
where $\Phi_i = (M_{ij}, B_i, \tilde{B}_i)$ and  $g_{i j^*}(\Phi)=\partial_{i}\partial_j^*K$ is the induced metric.

We argue that the vacuum and symmetry breaking pattern are entirely determined by the potential $V(\Phi_i)$ and do not depend on $g_{ij*}$. While the kinetic term can modify the expectation values of fields, it cannot alter the symmetry breaking pattern or change the vacuum energy. It is instructive to think of the two-derivative terms in analogy to general relativity, where the curvature can modify the notion of distance, but the vacuum energy is a gravitational invariant; thus, coordinate redefinitions of the scalar fields cannot affect its value. Therefore, the inclusion of $g_{ij*}$ would modify the values of the vacuum expectation value without affecting the depth of the potential, and thus a global minima would stay a global minima for any $g_{ij*}$.

Specifically, under a field redefinition of the form $\Phi^i \rightarrow R(\Phi^i) \Phi^i$, where $R(\Phi^i)$ is an invertible transformation, the location of the minimum in field space may shift, but the vacuum energy remains unchanged.  To further clarify why the symmetry breaking pattern is unaffected, we consider the mass matrix $\mathcal{M}$, which can be viewed as a complex symmetric bilinear form on the manifold described by the Kähler metric $g_{ij*}$. Under the transformation $R(\Phi^i)$, the mass matrix transforms as $R^\dagger \mathcal{M} R$. Since $R(\Phi^i)$ is invertible, the rank of the mass matrix remains unchanged, as this is just a similarity transformation. This invariance ensures that the Goldstone boson counting is unaffected: we can then diagonalize the potential $V$ into its mass basis before addressing the metric, preserving the same number of Goldstone modes.  

The only non-trivial consequence of the two-derivative terms comes from the restriction that the kinetic term remains positive. When the Lagrangian is of the form of Eq.~\eqref{eq:KählerMod}, the theory makes sense only if all the eigenvalues of the Kähler metric are positive. This, in turn, puts restrictions on the parameters of the theory. We consider these conditions in all of our scans, making sure to check the eigenvalues of the metric and demand positivity.

To highlight all these features, we can work out a simple toy model with one real scalar field and the Lagrangian given by
\begin{align}
\mathcal{L} = \frac{1}{2}(1-c \,\phi) \partial_\mu \phi \partial^\mu \phi - \lambda (\phi^2 - v^2)^2 \, .
\label{eq:toy}
\end{align}
This model, as seen as a perturbative low energy effective field theory, only makes sense when $c<1/v$. Therefore, while the potential is sufficient to compute the vacuum energy and the number of massless modes, there are non-trivial constraints that relate parameters in the potential to the parameters in the two derivative terms. Operators like the $c$ term in Eq.~\eqref{eq:toy} multiplicatively modify the mass of the massive states. 

To analyze the toy model of Eq.~\eqref{eq:toy}, one strategy is to first shift to a field that has no VEV: $\phi_0 = \phi - v$. One can then do a field redefinition of the form $\eta = \phi_0/\sqrt{1-c\,v}$. The terms with derivatives of $\eta$ in the Lagrangian are of the form 
\begin{equation}
   \mathcal{L} \supset \frac{1}{2}\partial_\mu \eta \partial^\mu \eta
    \left( 1+ \frac{c}{\sqrt{1-c \, v}}\eta \right) \, . 
    \label{eq:kin}
\end{equation}
The $\eta$ field then has a canonical kinetic term. The mass spectrum can be computed from the potential in terms of $\eta$, and the second term in Eq.~\eqref{eq:kin} is an interaction that can be ignored to find the spectrum at the tree-level.

An equivalent way to analyze the toy model is to use the following nonlinear field redefinition: 
\begin{equation}
    \psi= \frac{2}{3c}\left\{ \left(1-c \,\phi \right)^{3/2} - \left(1-c \, v  \right)^{3/2} 
    \right\}
\end{equation}
which puts the kinetic term in the canonical form. This redefinition does require that $c<1/v$ just as in the other method. The potential becomes more complicated, but one can verify that the vacuum energy is unchanged and that the mass of the field also remains the same.

\section{Wilson coefficients away from NDA}
\label{ap:notNDA}

In this work, we mainly focused on the low energy constants following NDA scaling. However, it is not necessarily true that in the perturbative regime, this hierarchy between the parameters is going to be preserved. To explore if there is any additional symmetry breaking pattern that occurs only outside of NDA scaling, we fix $m_{3/2}$ and
vary the size of the Wilson coefficients between 0 and $4\pi$. As above, we explore turning on only $c_B$ and also having all the Wilson coefficients nonzero but equally contributing. We fix the values of the superpotential parameters $\kappa$ and $\lambda$ using NDA.

\begin{figure}[h!]
 \resizebox{0.3285\linewidth}{!}{ \includegraphics{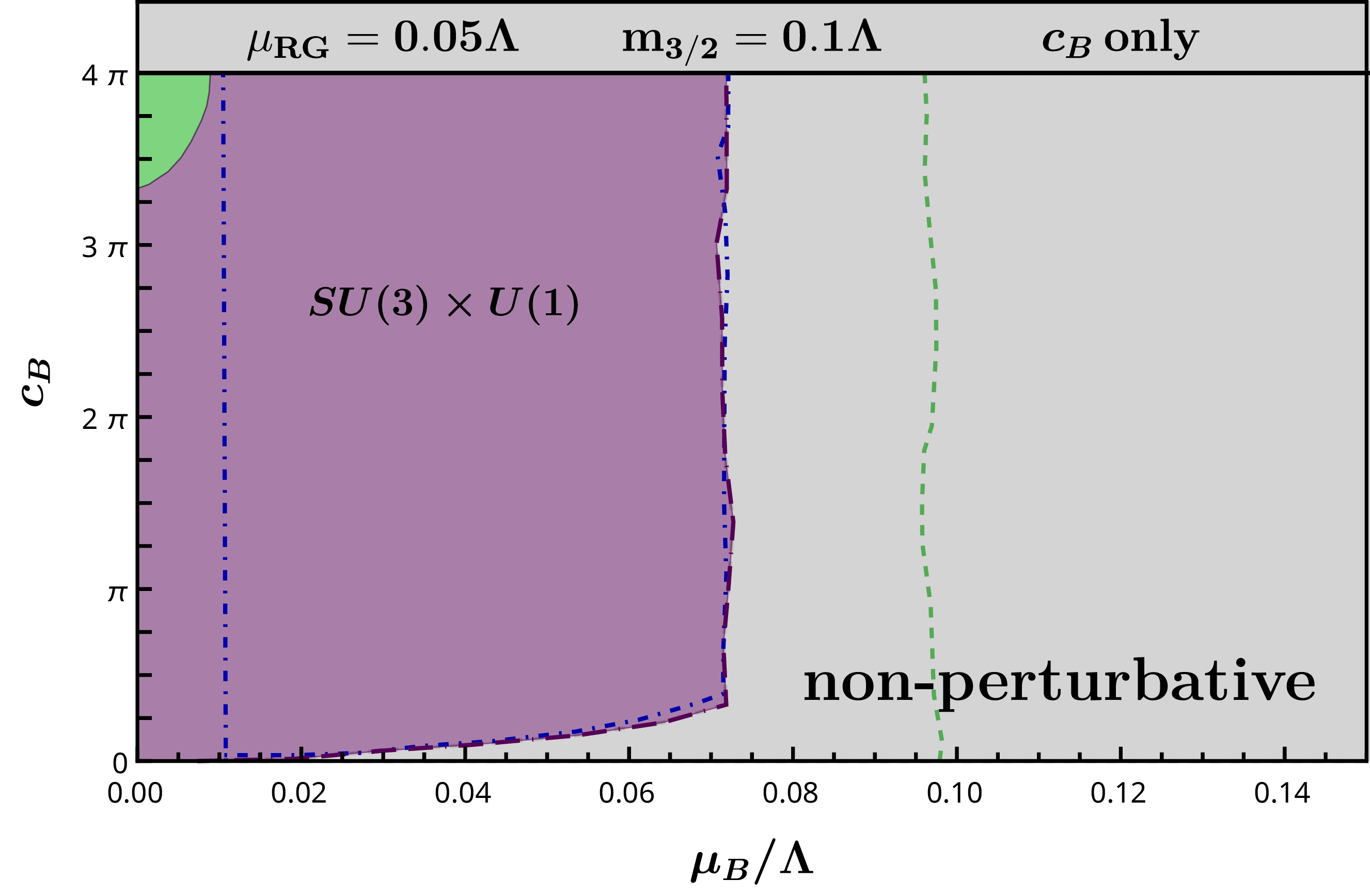}} 
 \resizebox{0.3285\linewidth}{!}{ \includegraphics{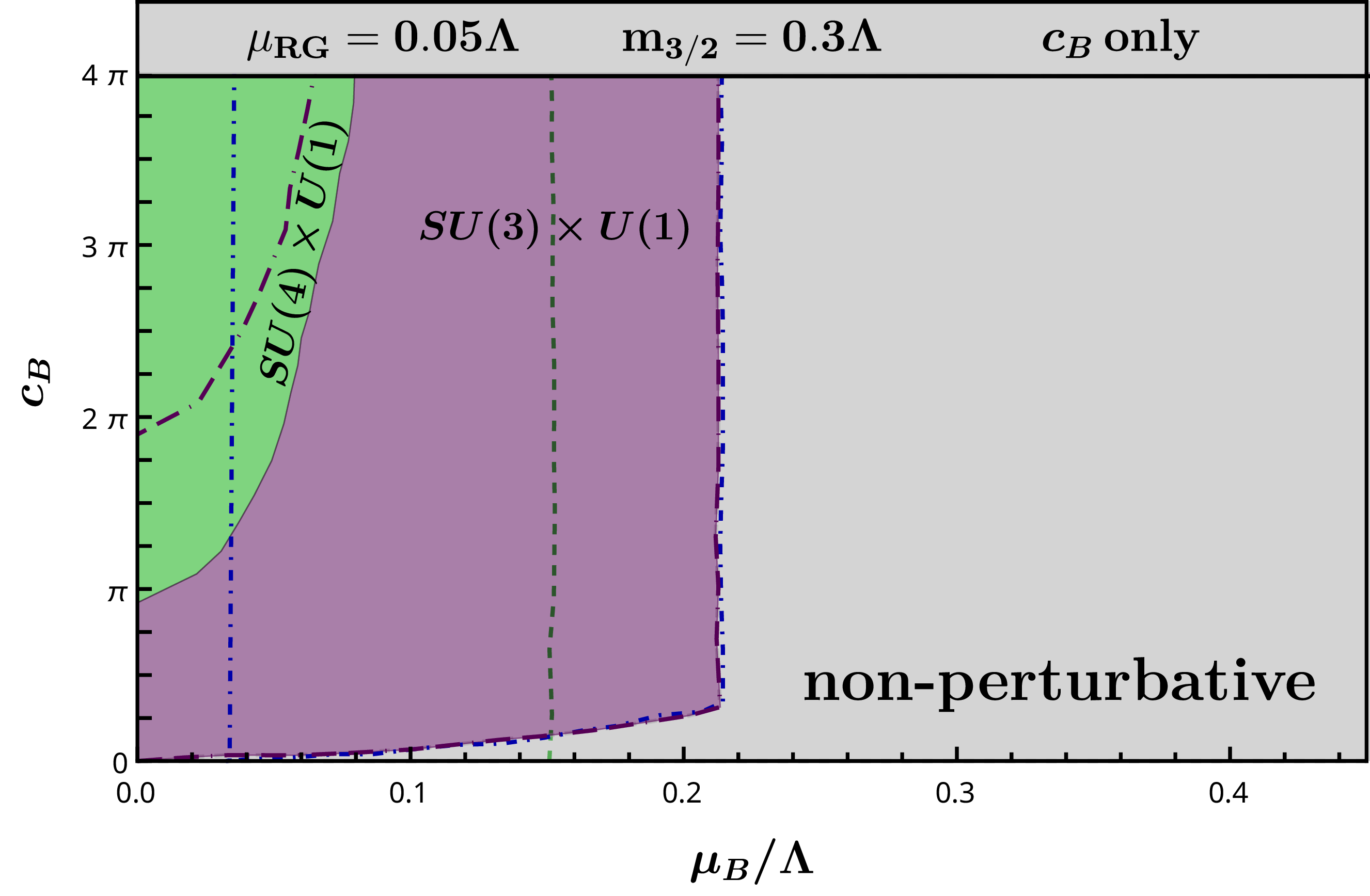}} 
 \resizebox{0.3285\linewidth}{!}{ \includegraphics{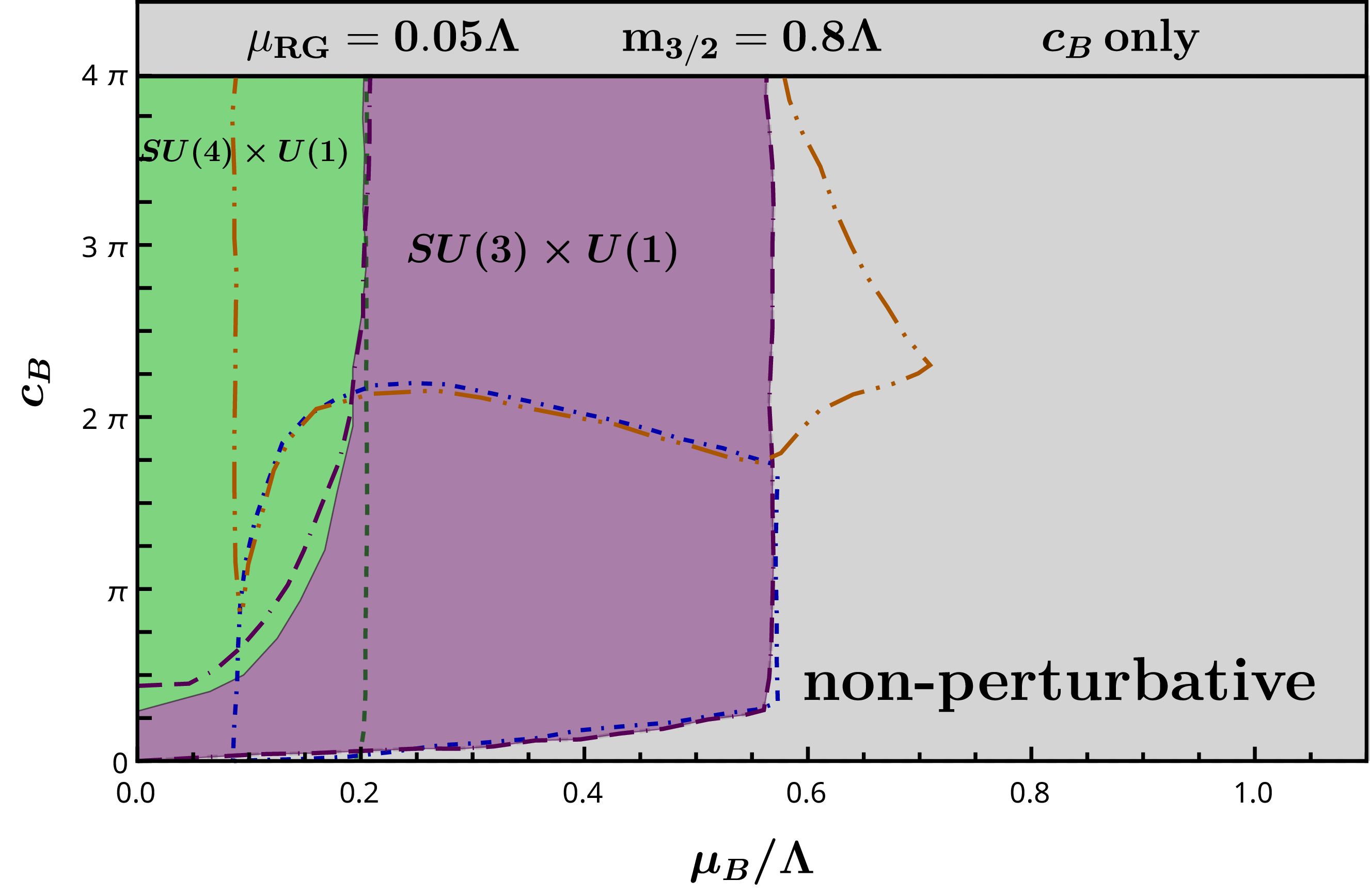}} 
 \resizebox{0.3285\linewidth}{!}{ \includegraphics{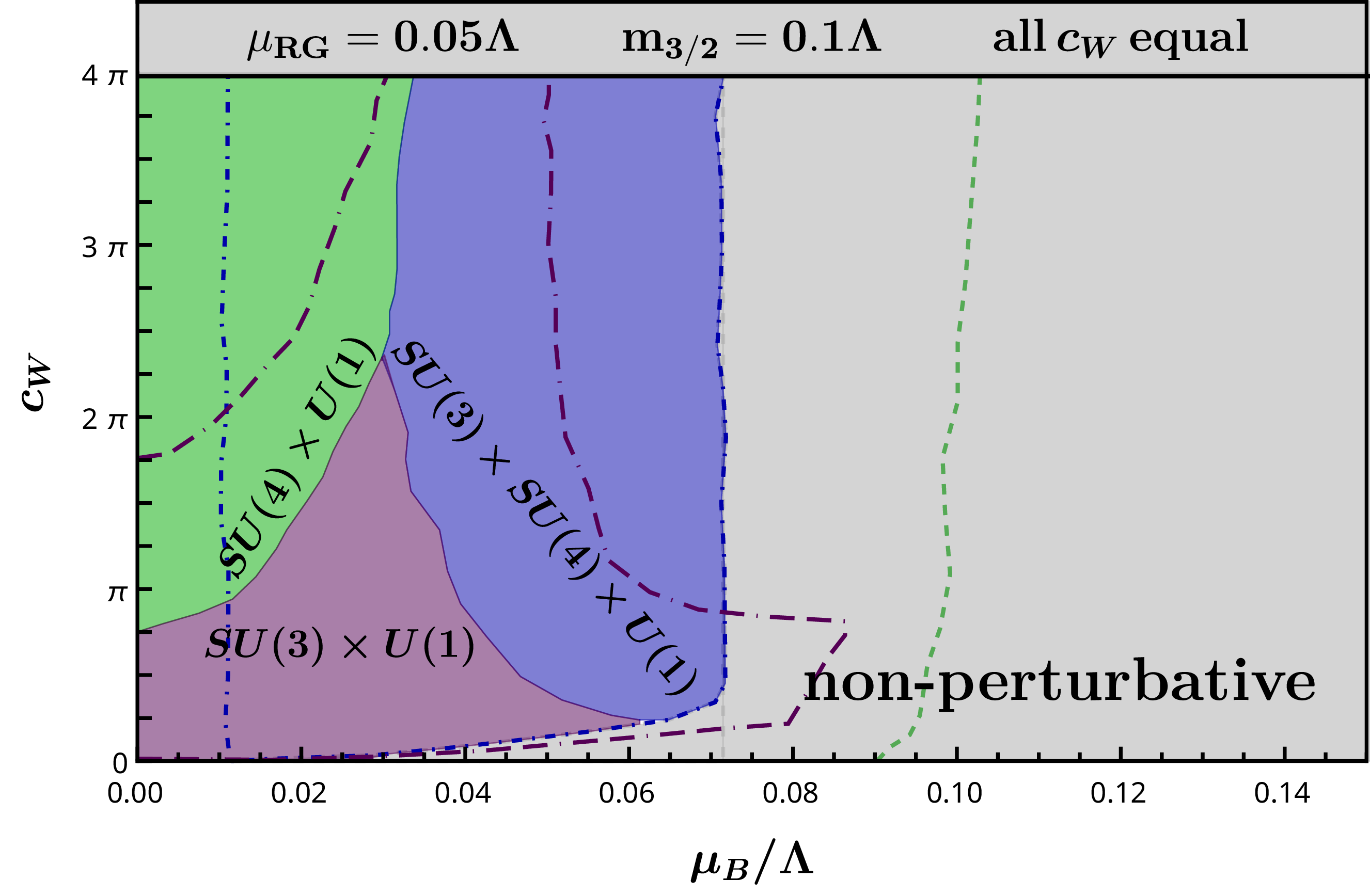}} 
 \resizebox{0.3285\linewidth}{!}{ \includegraphics{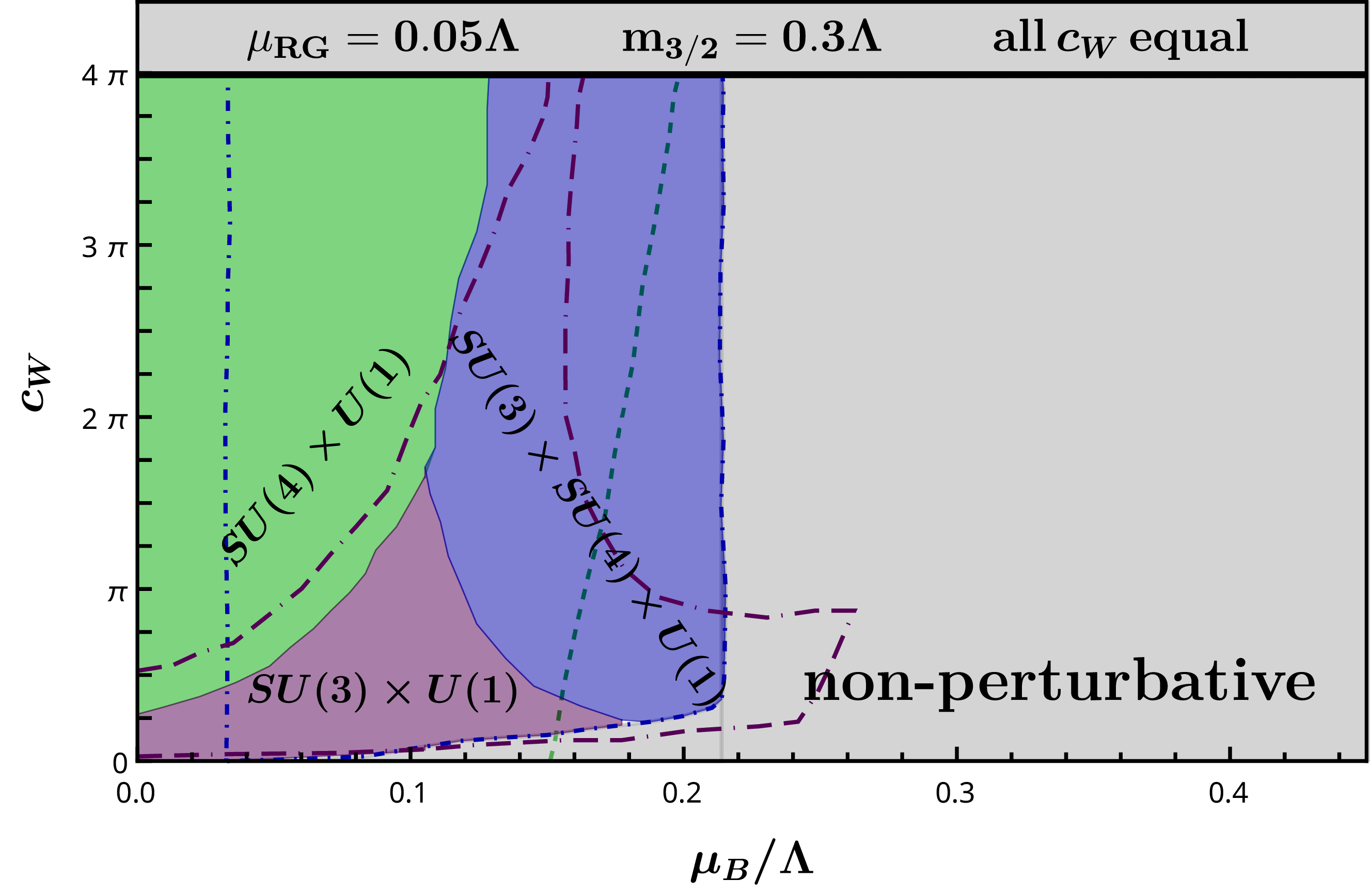}} 
 \resizebox{0.3285\linewidth}{!}{ \includegraphics{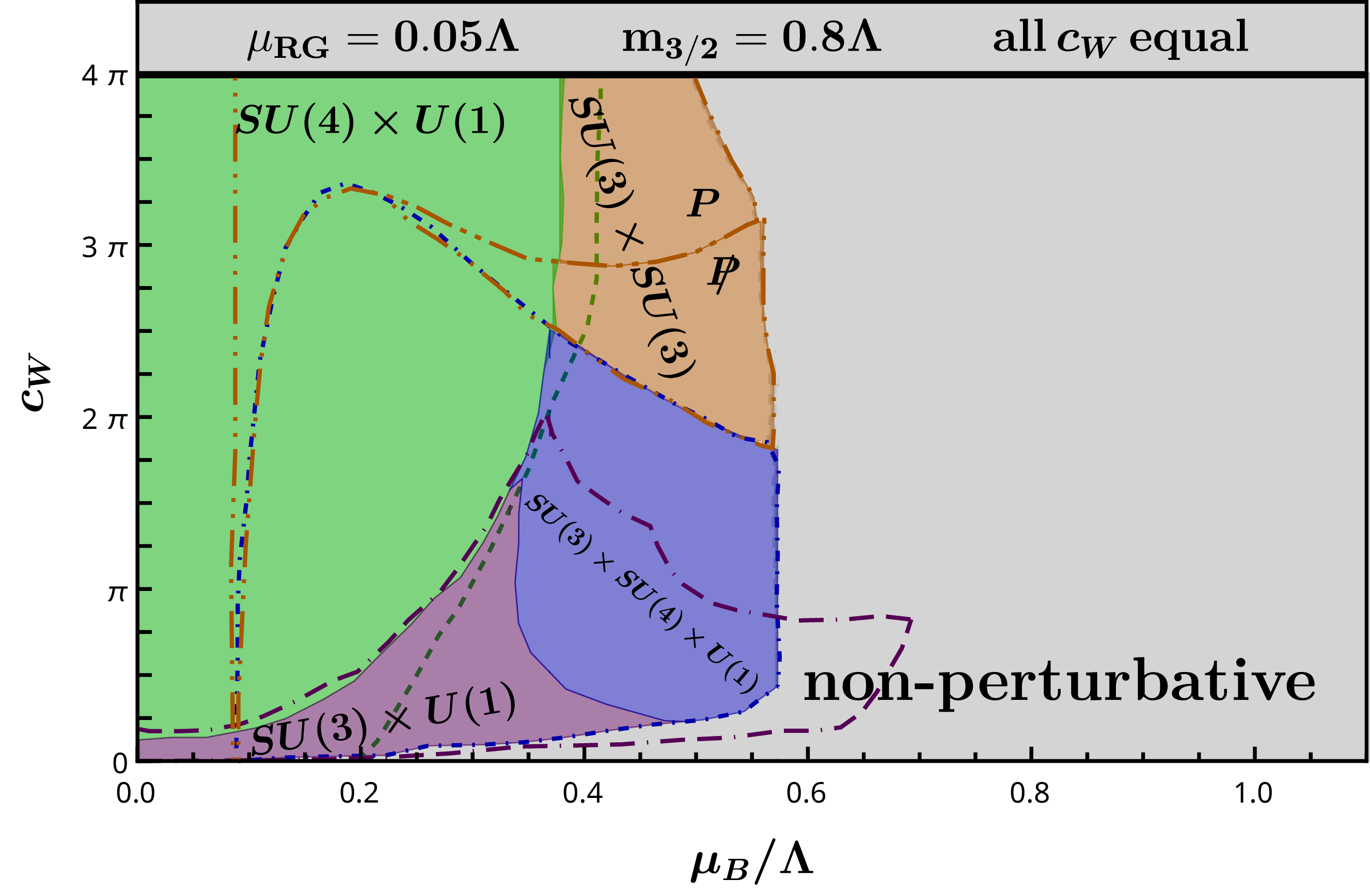}} 
\caption{Phase diagram in the plane of the Wilson coefficient $c_{i}$ vs baryon chemical potential $\mu_B$ for the couplings fixed by $\mu_{\rm RG}$ and different values of $m_{3/2}$. On the top row, we include only one nonzero Kähler coefficient, $c_B$, while on the bottom row, we include all dimension-six operators with the same overall size. The color and curve scheme is the same as in the other figures of the main article. One distinct feature in the bottom right plot is the presence of the $SU(3)\times SU(3)$ vacuum with the spontaneous breaking of parity or its preservation, mostly dependent on the value of $c_W$.}
\label{fig:FIX2}
\end{figure}

The results of our scan as a function of the Wilson coefficients and $\mu_B$ are shown in Fig.~\ref{fig:FIX2}. One can see that the phase depends on the value of the Wilson coefficients. However, again, one can see the overall trend that at larger values of $m_{3/2}$, the QCD-like vacuum transits to one of the exotic symmetry breaking patterns shown in the main article. Additionally, we highlight that no breaking pattern outside of the ones covered in the main article was found, even as we allow for parameter values outside of the NDA hierarchy.

\bibliographystyle{apsrev-title}
\bibliography{bibSUSYmu}

\begin{thebibliography}{53}
\expandafter\ifx\csname natexlab\endcsname\relax\def\natexlab#1{#1}\fi
\expandafter\ifx\csname bibnamefont\endcsname\relax
  \def\bibnamefont#1{#1}\fi
\expandafter\ifx\csname bibfnamefont\endcsname\relax
  \def\bibfnamefont#1{#1}\fi
\expandafter\ifx\csname citenamefont\endcsname\relax
  \def\citenamefont#1{#1}\fi
\expandafter\ifx\csname url\endcsname\relax
  \def\url#1{\texttt{#1}}\fi
\expandafter\ifx\csname urlprefix\endcsname\relax\def\urlprefix{URL }\fi
\providecommand{\bibinfo}[2]{#2}
\providecommand{\eprint}[2][]{\url{#2}}

\bibitem[{\citenamefont{Rajagopal and Wilczek}(2000)}]{Rajagopal:2000wf}
\bibinfo{author}{\bibfnamefont{K.}~\bibnamefont{Rajagopal}} \bibnamefont{and} \bibinfo{author}{\bibfnamefont{F.}~\bibnamefont{Wilczek}}, \emph{\bibinfo{title}{{The Condensed matter physics of QCD}}} (\bibinfo{year}{2000}), pp. \bibinfo{pages}{2061--2151}, \eprint{hep-ph/0011333}.

\bibitem[{\citenamefont{Alford}(2001)}]{Alford:2001dt}
\bibinfo{author}{\bibfnamefont{M.~G.} \bibnamefont{Alford}}, ``{Color superconducting quark matter},'' \bibinfo{journal}{Ann. Rev. Nucl. Part. Sci.} \textbf{\bibinfo{volume}{51}}, \bibinfo{pages}{131} (\bibinfo{year}{2001}), \eprint{hep-ph/0102047}.

\bibitem[{\citenamefont{Sch\"afer}(2000)}]{Schafer:2000tw}
\bibinfo{author}{\bibfnamefont{T.}~\bibnamefont{Sch\"afer}}, ``{Quark hadron continuity in QCD with one flavor},'' \bibinfo{journal}{Phys. Rev. D} \textbf{\bibinfo{volume}{62}}, \bibinfo{pages}{094007} (\bibinfo{year}{2000}), \eprint{hep-ph/0006034}.

\bibitem[{\citenamefont{Alford et~al.}(2001)\citenamefont{Alford, Bowers, and Rajagopal}}]{Alford:2000ze}
\bibinfo{author}{\bibfnamefont{M.~G.} \bibnamefont{Alford}}, \bibinfo{author}{\bibfnamefont{J.~A.} \bibnamefont{Bowers}}, \bibnamefont{and} \bibinfo{author}{\bibfnamefont{K.}~\bibnamefont{Rajagopal}}, ``{Crystalline color superconductivity},'' \bibinfo{journal}{Phys. Rev. D} \textbf{\bibinfo{volume}{63}}, \bibinfo{pages}{074016} (\bibinfo{year}{2001}), \eprint{hep-ph/0008208}.

\bibitem[{\citenamefont{Vovchenko et~al.}(2019)\citenamefont{Vovchenko, Steinheimer, Philipsen, Pasztor, Fodor, Katz, and Stoecker}}]{Vovchenko:2018zgt}
\bibinfo{author}{\bibfnamefont{V.}~\bibnamefont{Vovchenko}}, \bibinfo{author}{\bibfnamefont{J.}~\bibnamefont{Steinheimer}}, \bibinfo{author}{\bibfnamefont{O.}~\bibnamefont{Philipsen}}, \bibinfo{author}{\bibfnamefont{A.}~\bibnamefont{Pasztor}}, \bibinfo{author}{\bibfnamefont{Z.}~\bibnamefont{Fodor}}, \bibinfo{author}{\bibfnamefont{S.~D.} \bibnamefont{Katz}}, \bibnamefont{and} \bibinfo{author}{\bibfnamefont{H.}~\bibnamefont{Stoecker}}, ``{Lattice-based QCD equation of state at finite baryon density: Cluster Expansion Model},'' \bibinfo{journal}{Nucl. Phys. A} \textbf{\bibinfo{volume}{982}}, \bibinfo{pages}{859} (\bibinfo{year}{2019}), \eprint{1807.06472}.

\bibitem[{\citenamefont{D'Elia and Lombardo}(2003)}]{DElia:2002tig}
\bibinfo{author}{\bibfnamefont{M.}~\bibnamefont{D'Elia}} \bibnamefont{and} \bibinfo{author}{\bibfnamefont{M.-P.} \bibnamefont{Lombardo}}, ``{Finite density QCD via imaginary chemical potential},'' \bibinfo{journal}{Phys. Rev. D} \textbf{\bibinfo{volume}{67}}, \bibinfo{pages}{014505} (\bibinfo{year}{2003}), \eprint{hep-lat/0209146}.

\bibitem[{\citenamefont{de~Forcrand and Kratochvila}(2006)}]{deForcrand:2006ec}
\bibinfo{author}{\bibfnamefont{P.}~\bibnamefont{de~Forcrand}} \bibnamefont{and} \bibinfo{author}{\bibfnamefont{S.}~\bibnamefont{Kratochvila}}, ``{Finite density QCD with a canonical approach},'' \bibinfo{journal}{Nucl. Phys. B Proc. Suppl.} \textbf{\bibinfo{volume}{153}}, \bibinfo{pages}{62} (\bibinfo{year}{2006}), \eprint{hep-lat/0602024}.

\bibitem[{\citenamefont{Baym et~al.}(2018)\citenamefont{Baym, Hatsuda, Kojo, Powell, Song, and Takatsuka}}]{Baym:2017whm}
\bibinfo{author}{\bibfnamefont{G.}~\bibnamefont{Baym}}, \bibinfo{author}{\bibfnamefont{T.}~\bibnamefont{Hatsuda}}, \bibinfo{author}{\bibfnamefont{T.}~\bibnamefont{Kojo}}, \bibinfo{author}{\bibfnamefont{P.~D.} \bibnamefont{Powell}}, \bibinfo{author}{\bibfnamefont{Y.}~\bibnamefont{Song}}, \bibnamefont{and} \bibinfo{author}{\bibfnamefont{T.}~\bibnamefont{Takatsuka}}, ``{From hadrons to quarks in neutron stars: a review},'' \bibinfo{journal}{Rept. Prog. Phys.} \textbf{\bibinfo{volume}{81}}, \bibinfo{pages}{056902} (\bibinfo{year}{2018}), \eprint{1707.04966}.

\bibitem[{\citenamefont{Wu et~al.}(2021)\citenamefont{Wu, Ping, and Zong}}]{Wu:2020qgr}
\bibinfo{author}{\bibfnamefont{Z.-Q.} \bibnamefont{Wu}}, \bibinfo{author}{\bibfnamefont{J.-L.} \bibnamefont{Ping}}, \bibnamefont{and} \bibinfo{author}{\bibfnamefont{H.-S.} \bibnamefont{Zong}}, ``{QCD phase diagram at finite isospin and baryon chemical potentials with the self-consistent mean field approximation},'' \bibinfo{journal}{Chin. Phys. C} \textbf{\bibinfo{volume}{45}}, \bibinfo{pages}{064102} (\bibinfo{year}{2021}), \eprint{2009.13070}.

\bibitem[{\citenamefont{Seiberg}(1994)}]{Seiberg:1994bz}
\bibinfo{author}{\bibfnamefont{N.}~\bibnamefont{Seiberg}}, ``{Exact results on the space of vacua of four-dimensional SUSY gauge theories},'' \bibinfo{journal}{Phys. Rev. D} \textbf{\bibinfo{volume}{49}}, \bibinfo{pages}{6857} (\bibinfo{year}{1994}), \eprint{hep-th/9402044}.

\bibitem[{\citenamefont{Seiberg}(1995)}]{Seiberg:1994pq}
\bibinfo{author}{\bibfnamefont{N.}~\bibnamefont{Seiberg}}, ``{Electric - magnetic duality in supersymmetric nonAbelian gauge theories},'' \bibinfo{journal}{Nucl. Phys. B} \textbf{\bibinfo{volume}{435}}, \bibinfo{pages}{129} (\bibinfo{year}{1995}), \eprint{hep-th/9411149}.

\bibitem[{\citenamefont{Harnik et~al.}(2004)\citenamefont{Harnik, Larson, and Murayama}}]{Harnik:2003ke}
\bibinfo{author}{\bibfnamefont{R.}~\bibnamefont{Harnik}}, \bibinfo{author}{\bibfnamefont{D.~T.} \bibnamefont{Larson}}, \bibnamefont{and} \bibinfo{author}{\bibfnamefont{H.}~\bibnamefont{Murayama}}, ``{Supersymmetric color superconductivity},'' \bibinfo{journal}{JHEP} \textbf{\bibinfo{volume}{03}}, \bibinfo{pages}{049} (\bibinfo{year}{2004}), \eprint{hep-ph/0309224}.

\bibitem[{\citenamefont{Randall and Sundrum}(1999)}]{Randall:1998uk}
\bibinfo{author}{\bibfnamefont{L.}~\bibnamefont{Randall}} \bibnamefont{and} \bibinfo{author}{\bibfnamefont{R.}~\bibnamefont{Sundrum}}, ``{Out of this world supersymmetry breaking},'' \bibinfo{journal}{Nucl. Phys. B} \textbf{\bibinfo{volume}{557}}, \bibinfo{pages}{79} (\bibinfo{year}{1999}), \eprint{hep-th/9810155}.

\bibitem[{\citenamefont{Giudice et~al.}(1998)\citenamefont{Giudice, Luty, Murayama, and Rattazzi}}]{Giudice:1998xp}
\bibinfo{author}{\bibfnamefont{G.~F.} \bibnamefont{Giudice}}, \bibinfo{author}{\bibfnamefont{M.~A.} \bibnamefont{Luty}}, \bibinfo{author}{\bibfnamefont{H.}~\bibnamefont{Murayama}}, \bibnamefont{and} \bibinfo{author}{\bibfnamefont{R.}~\bibnamefont{Rattazzi}}, ``{Gaugino mass without singlets},'' \bibinfo{journal}{JHEP} \textbf{\bibinfo{volume}{12}}, \bibinfo{pages}{027} (\bibinfo{year}{1998}), \eprint{hep-ph/9810442}.

\bibitem[{\citenamefont{Pomarol and Rattazzi}(1999)}]{Pomarol:1999ie}
\bibinfo{author}{\bibfnamefont{A.}~\bibnamefont{Pomarol}} \bibnamefont{and} \bibinfo{author}{\bibfnamefont{R.}~\bibnamefont{Rattazzi}}, ``{Sparticle masses from the superconformal anomaly},'' \bibinfo{journal}{JHEP} \textbf{\bibinfo{volume}{05}}, \bibinfo{pages}{013} (\bibinfo{year}{1999}), \eprint{hep-ph/9903448}.

\bibitem[{\citenamefont{Jung and Lee}(2009)}]{Jung:2009dg}
\bibinfo{author}{\bibfnamefont{D.-W.} \bibnamefont{Jung}} \bibnamefont{and} \bibinfo{author}{\bibfnamefont{J.~Y.} \bibnamefont{Lee}}, ``{Anomaly-Mediated Supersymmetry Breaking Demystified},'' \bibinfo{journal}{JHEP} \textbf{\bibinfo{volume}{03}}, \bibinfo{pages}{123} (\bibinfo{year}{2009}), \eprint{0902.0464}.

\bibitem[{\citenamefont{Murayama}(2021)}]{Murayama:2021xfj}
\bibinfo{author}{\bibfnamefont{H.}~\bibnamefont{Murayama}}, ``{Some Exact Results in QCD-like Theories},'' \bibinfo{journal}{Phys. Rev. Lett.} \textbf{\bibinfo{volume}{126}}, \bibinfo{pages}{251601} (\bibinfo{year}{2021}), \eprint{2104.01179}.

\bibitem[{\citenamefont{Cs\'aki et~al.}(2022{\natexlab{a}})\citenamefont{Cs\'aki, Murayama, and Telem}}]{Csaki:2021aqv}
\bibinfo{author}{\bibfnamefont{C.}~\bibnamefont{Cs\'aki}}, \bibinfo{author}{\bibfnamefont{H.}~\bibnamefont{Murayama}}, \bibnamefont{and} \bibinfo{author}{\bibfnamefont{O.}~\bibnamefont{Telem}}, ``{More exact results on chiral gauge theories: The case of the symmetric tensor},'' \bibinfo{journal}{Phys. Rev. D} \textbf{\bibinfo{volume}{105}}, \bibinfo{pages}{045007} (\bibinfo{year}{2022}{\natexlab{a}}), \eprint{2105.03444}.

\bibitem[{\citenamefont{Cs\'aki et~al.}(2021)\citenamefont{Cs\'aki, Gomes, Murayama, and Telem}}]{Csaki:2021jax}
\bibinfo{author}{\bibfnamefont{C.}~\bibnamefont{Cs\'aki}}, \bibinfo{author}{\bibfnamefont{A.}~\bibnamefont{Gomes}}, \bibinfo{author}{\bibfnamefont{H.}~\bibnamefont{Murayama}}, \bibnamefont{and} \bibinfo{author}{\bibfnamefont{O.}~\bibnamefont{Telem}}, ``{Demonstration of Confinement and Chiral Symmetry Breaking in SO(Nc) Gauge Theories},'' \bibinfo{journal}{Phys. Rev. Lett.} \textbf{\bibinfo{volume}{127}}, \bibinfo{pages}{251602} (\bibinfo{year}{2021}), \eprint{2106.10288}.

\bibitem[{\citenamefont{Bai and Stolarski}(2022)}]{Bai:2021tgl}
\bibinfo{author}{\bibfnamefont{Y.}~\bibnamefont{Bai}} \bibnamefont{and} \bibinfo{author}{\bibfnamefont{D.}~\bibnamefont{Stolarski}}, ``{Phases of confining SU(5) chiral gauge theory with three generations},'' \bibinfo{journal}{JHEP} \textbf{\bibinfo{volume}{03}}, \bibinfo{pages}{113} (\bibinfo{year}{2022}), \eprint{2111.11214}.

\bibitem[{\citenamefont{Luzio and Xu}(2022)}]{Luzio:2022ccn}
\bibinfo{author}{\bibfnamefont{A.}~\bibnamefont{Luzio}} \bibnamefont{and} \bibinfo{author}{\bibfnamefont{L.-X.} \bibnamefont{Xu}}, ``{On the derivation of chiral symmetry breaking in QCD-like theories and S-confining theories},'' \bibinfo{journal}{JHEP} \textbf{\bibinfo{volume}{08}}, \bibinfo{pages}{016} (\bibinfo{year}{2022}), \eprint{2202.01239}.

\bibitem[{\citenamefont{Kondo et~al.}(2022)\citenamefont{Kondo, Murayama, and Sylber}}]{Kondo:2022lvu}
\bibinfo{author}{\bibfnamefont{D.}~\bibnamefont{Kondo}}, \bibinfo{author}{\bibfnamefont{H.}~\bibnamefont{Murayama}}, \bibnamefont{and} \bibinfo{author}{\bibfnamefont{C.}~\bibnamefont{Sylber}}, ``{Dynamics of Simplest Chiral Gauge Theories},''  (\bibinfo{year}{2022}), \eprint{2209.09287}.

\bibitem[{\citenamefont{Cs\'aki et~al.}(2022{\natexlab{b}})\citenamefont{Cs\'aki, Gomes, Murayama, Noether, Varier, and Telem}}]{Csaki:2022cyg}
\bibinfo{author}{\bibfnamefont{C.}~\bibnamefont{Cs\'aki}}, \bibinfo{author}{\bibfnamefont{A.}~\bibnamefont{Gomes}}, \bibinfo{author}{\bibfnamefont{H.}~\bibnamefont{Murayama}}, \bibinfo{author}{\bibfnamefont{B.}~\bibnamefont{Noether}}, \bibinfo{author}{\bibfnamefont{D.~R.} \bibnamefont{Varier}}, \bibnamefont{and} \bibinfo{author}{\bibfnamefont{O.}~\bibnamefont{Telem}}, ``{A Guide to AMSB QCD},''  (\bibinfo{year}{2022}{\natexlab{b}}), \eprint{2212.03260}.

\bibitem[{\citenamefont{de~Lima and Stolarski}(2023)}]{deLima:2023ebw}
\bibinfo{author}{\bibfnamefont{C.~H.} \bibnamefont{de~Lima}} \bibnamefont{and} \bibinfo{author}{\bibfnamefont{D.}~\bibnamefont{Stolarski}}, ``{On s-confining SUSY-QCD with anomaly mediation},'' \bibinfo{journal}{JHEP} \textbf{\bibinfo{volume}{10}}, \bibinfo{pages}{020} (\bibinfo{year}{2023}), \eprint{2307.13154}.

\bibitem[{\citenamefont{Cs\'aki et~al.}(2024)\citenamefont{Cs\'aki, Ruhdorfer, and Youn}}]{Csaki:2024lvk}
\bibinfo{author}{\bibfnamefont{C.}~\bibnamefont{Cs\'aki}}, \bibinfo{author}{\bibfnamefont{M.}~\bibnamefont{Ruhdorfer}}, \bibnamefont{and} \bibinfo{author}{\bibfnamefont{T.}~\bibnamefont{Youn}}, ``{Spontaneous CP breaking in a QCD-like theory},'' \bibinfo{journal}{JHEP} \textbf{\bibinfo{volume}{12}}, \bibinfo{pages}{066} (\bibinfo{year}{2024}), \eprint{2407.06252}.

\bibitem[{\citenamefont{Leedom et~al.}(2025)\citenamefont{Leedom, Murayama, Singh, Suter, and Wong}}]{Leedom:2025mcg}
\bibinfo{author}{\bibfnamefont{J.~M.} \bibnamefont{Leedom}}, \bibinfo{author}{\bibfnamefont{H.}~\bibnamefont{Murayama}}, \bibinfo{author}{\bibfnamefont{G.}~\bibnamefont{Singh}}, \bibinfo{author}{\bibfnamefont{B.}~\bibnamefont{Suter}}, \bibnamefont{and} \bibinfo{author}{\bibfnamefont{J.}~\bibnamefont{Wong}}, ``{Exact Results in Chiral Gauge Theories with Flavor},''  (\bibinfo{year}{2025}), \eprint{2503.08772}.

\bibitem[{\citenamefont{Goh et~al.}(2025)\citenamefont{Goh, Murayama, Singh, Suter, and Wong}}]{Goh:2025oes}
\bibinfo{author}{\bibfnamefont{A.}~\bibnamefont{Goh}}, \bibinfo{author}{\bibfnamefont{H.}~\bibnamefont{Murayama}}, \bibinfo{author}{\bibfnamefont{G.}~\bibnamefont{Singh}}, \bibinfo{author}{\bibfnamefont{B.}~\bibnamefont{Suter}}, \bibnamefont{and} \bibinfo{author}{\bibfnamefont{J.}~\bibnamefont{Wong}}, ``{Dynamics of $E_6$ Chiral Gauge Theories},''  (\bibinfo{year}{2025}), \eprint{2505.07931}.

\bibitem[{\citenamefont{Kondo et~al.}(2025)\citenamefont{Kondo, Murayama, and Noether}}]{Kondo:2025njf}
\bibinfo{author}{\bibfnamefont{D.}~\bibnamefont{Kondo}}, \bibinfo{author}{\bibfnamefont{H.}~\bibnamefont{Murayama}}, \bibnamefont{and} \bibinfo{author}{\bibfnamefont{B.}~\bibnamefont{Noether}}, ``{Near-SUSY to Non-SUSY Crossover},''  (\bibinfo{year}{2025}), \eprint{2505.18138}.

\bibitem[{\citenamefont{Palkanoglou et~al.}(2025)\citenamefont{Palkanoglou, Stuck, and Gezerlis}}]{Palkanoglou:2024epd}
\bibinfo{author}{\bibfnamefont{G.}~\bibnamefont{Palkanoglou}}, \bibinfo{author}{\bibfnamefont{M.}~\bibnamefont{Stuck}}, \bibnamefont{and} \bibinfo{author}{\bibfnamefont{A.}~\bibnamefont{Gezerlis}}, ``{Spin-Triplet Pairing in Heavy Nuclei Is Stable against Deformation},'' \bibinfo{journal}{Phys. Rev. Lett.} \textbf{\bibinfo{volume}{134}}, \bibinfo{pages}{032501} (\bibinfo{year}{2025}), \eprint{2402.13313}.

\bibitem[{\citenamefont{Palkanoglou and Gezerlis}(2025)}]{Palkanoglou:2025guj}
\bibinfo{author}{\bibfnamefont{G.}~\bibnamefont{Palkanoglou}} \bibnamefont{and} \bibinfo{author}{\bibfnamefont{A.}~\bibnamefont{Gezerlis}}, ``{Symmetry properties of pair correlations in heavy deformed nuclei},''  (\bibinfo{year}{2025}), \eprint{2505.08879}.

\bibitem[{\citenamefont{Frauendorf and Macchiavelli}(2014)}]{Frauendorf:2014mja}
\bibinfo{author}{\bibfnamefont{S.}~\bibnamefont{Frauendorf}} \bibnamefont{and} \bibinfo{author}{\bibfnamefont{A.~O.} \bibnamefont{Macchiavelli}}, ``{Overview of neutron\textendash{}proton pairing},'' \bibinfo{journal}{Prog. Part. Nucl. Phys.} \textbf{\bibinfo{volume}{78}}, \bibinfo{pages}{24} (\bibinfo{year}{2014}), \eprint{1405.1652}.

\bibitem[{\citenamefont{Dean and Hjorth-Jensen}(2003)}]{Dean:2002zx}
\bibinfo{author}{\bibfnamefont{D.~J.} \bibnamefont{Dean}} \bibnamefont{and} \bibinfo{author}{\bibfnamefont{M.}~\bibnamefont{Hjorth-Jensen}}, ``{Pairing in nuclear systems: From neutron stars to finite nuclei},'' \bibinfo{journal}{Rev. Mod. Phys.} \textbf{\bibinfo{volume}{75}}, \bibinfo{pages}{607} (\bibinfo{year}{2003}), \eprint{nucl-th/0210033}.

\bibitem[{\citenamefont{Sedrakian and Clark}(2019)}]{Sedrakian:2018ydt}
\bibinfo{author}{\bibfnamefont{A.}~\bibnamefont{Sedrakian}} \bibnamefont{and} \bibinfo{author}{\bibfnamefont{J.~W.} \bibnamefont{Clark}}, ``{Superfluidity in nuclear systems and neutron stars},'' \bibinfo{journal}{Eur. Phys. J. A} \textbf{\bibinfo{volume}{55}}, \bibinfo{pages}{167} (\bibinfo{year}{2019}), \eprint{1802.00017}.

\bibitem[{\citenamefont{Bertsch and Luo}(2010)}]{Bertsch:2009xz}
\bibinfo{author}{\bibfnamefont{G.~F.} \bibnamefont{Bertsch}} \bibnamefont{and} \bibinfo{author}{\bibfnamefont{Y.}~\bibnamefont{Luo}}, ``{Spin-triplet pairing in large nuclei},'' \bibinfo{journal}{Phys. Rev. C} \textbf{\bibinfo{volume}{81}}, \bibinfo{pages}{064320} (\bibinfo{year}{2010}), \eprint{0912.2533}.

\bibitem[{\citenamefont{Bai et~al.}(2025)\citenamefont{Bai, de~Lima, and Stolarski}}]{tempAMSB}
\bibinfo{author}{\bibfnamefont{Y.}~\bibnamefont{Bai}}, \bibinfo{author}{\bibfnamefont{C.~H.} \bibnamefont{de~Lima}}, \bibnamefont{and} \bibinfo{author}{\bibfnamefont{D.}~\bibnamefont{Stolarski}} \bibinfo{journal}{, to appear}  (\bibinfo{year}{2025}).

\bibitem[{\citenamefont{Cherman et~al.}(2014)\citenamefont{Cherman, Grozdanov, and Hardy}}]{Cherman:2013rla}
\bibinfo{author}{\bibfnamefont{A.}~\bibnamefont{Cherman}}, \bibinfo{author}{\bibfnamefont{S.}~\bibnamefont{Grozdanov}}, \bibnamefont{and} \bibinfo{author}{\bibfnamefont{E.}~\bibnamefont{Hardy}}, ``{Searching for Fermi Surfaces in Super-QED},'' \bibinfo{journal}{JHEP} \textbf{\bibinfo{volume}{06}}, \bibinfo{pages}{046} (\bibinfo{year}{2014}), \eprint{1308.0335}.

\bibitem[{\citenamefont{Coleman and Weinberg}(1973)}]{PhysRevD.7.1888}
\bibinfo{author}{\bibfnamefont{S.}~\bibnamefont{Coleman}} \bibnamefont{and} \bibinfo{author}{\bibfnamefont{E.}~\bibnamefont{Weinberg}}, ``Radiative corrections as the origin of spontaneous symmetry breaking,'' \bibinfo{journal}{Phys. Rev. D} \textbf{\bibinfo{volume}{7}}, \bibinfo{pages}{1888} (\bibinfo{year}{1973}), \urlprefix\url{https://link.aps.org/doi/10.1103/PhysRevD.7.1888}.

\bibitem[{\citenamefont{Jackiw}(1974)}]{PhysRevD.9.1686}
\bibinfo{author}{\bibfnamefont{R.}~\bibnamefont{Jackiw}}, ``Functional evaluation of the effective potential,'' \bibinfo{journal}{Phys. Rev. D} \textbf{\bibinfo{volume}{9}}, \bibinfo{pages}{1686} (\bibinfo{year}{1974}), \urlprefix\url{https://link.aps.org/doi/10.1103/PhysRevD.9.1686}.

\bibitem[{\citenamefont{Martin}(2002)}]{Martin:2001vx}
\bibinfo{author}{\bibfnamefont{S.~P.} \bibnamefont{Martin}}, ``{Two Loop Effective Potential for a General Renormalizable Theory and Softly Broken Supersymmetry},'' \bibinfo{journal}{Phys. Rev. D} \textbf{\bibinfo{volume}{65}}, \bibinfo{pages}{116003} (\bibinfo{year}{2002}), \eprint{hep-ph/0111209}.

\bibitem[{\citenamefont{Martin}(2017)}]{Martin:2017lqn}
\bibinfo{author}{\bibfnamefont{S.~P.} \bibnamefont{Martin}}, ``{Effective potential at three loops},'' \bibinfo{journal}{Phys. Rev. D} \textbf{\bibinfo{volume}{96}}, \bibinfo{pages}{096005} (\bibinfo{year}{2017}), \eprint{1709.02397}.

\bibitem[{\citenamefont{Martin}(2024)}]{Martin:2024qmi}
\bibinfo{author}{\bibfnamefont{S.~P.} \bibnamefont{Martin}}, ``{Effective K\"ahler and auxiliary field potentials for chiral superfield models at three loops},''  (\bibinfo{year}{2024}), \eprint{2408.04589}.

\bibitem[{\citenamefont{Pickering and West}(1996)}]{Pickering:1996he}
\bibinfo{author}{\bibfnamefont{A.}~\bibnamefont{Pickering}} \bibnamefont{and} \bibinfo{author}{\bibfnamefont{P.~C.} \bibnamefont{West}}, ``{The One loop effective superpotential and nonholomorphicity},'' \bibinfo{journal}{Phys. Lett. B} \textbf{\bibinfo{volume}{383}}, \bibinfo{pages}{54} (\bibinfo{year}{1996}), \eprint{hep-th/9604147}.

\bibitem[{\citenamefont{Grisaru et~al.}(1996)\citenamefont{Grisaru, Rocek, and von Unge}}]{Grisaru:1996ve}
\bibinfo{author}{\bibfnamefont{M.~T.} \bibnamefont{Grisaru}}, \bibinfo{author}{\bibfnamefont{M.}~\bibnamefont{Rocek}}, \bibnamefont{and} \bibinfo{author}{\bibfnamefont{R.}~\bibnamefont{von Unge}}, ``{Effective Kahler potentials},'' \bibinfo{journal}{Phys. Lett. B} \textbf{\bibinfo{volume}{383}}, \bibinfo{pages}{415} (\bibinfo{year}{1996}), \eprint{hep-th/9605149}.

\bibitem[{\citenamefont{Brignole}(2000)}]{Brignole:2000kg}
\bibinfo{author}{\bibfnamefont{A.}~\bibnamefont{Brignole}}, ``{One loop Kahler potential in non renormalizable theories},'' \bibinfo{journal}{Nucl. Phys. B} \textbf{\bibinfo{volume}{579}}, \bibinfo{pages}{101} (\bibinfo{year}{2000}), \eprint{hep-th/0001121}.

\bibitem[{\citenamefont{Groot~Nibbelink and Nyawelo}(2006)}]{GrootNibbelink:2005nez}
\bibinfo{author}{\bibfnamefont{S.}~\bibnamefont{Groot~Nibbelink}} \bibnamefont{and} \bibinfo{author}{\bibfnamefont{T.~S.} \bibnamefont{Nyawelo}}, ``{Two Loop effective Kahler potential of (non-)renormalizable supersymmetric models},'' \bibinfo{journal}{JHEP} \textbf{\bibinfo{volume}{01}}, \bibinfo{pages}{034} (\bibinfo{year}{2006}), \eprint{hep-th/0511004}.

\bibitem[{\citenamefont{Kuzenko and Tyler}(2014)}]{Kuzenko:2014ypa}
\bibinfo{author}{\bibfnamefont{S.~M.} \bibnamefont{Kuzenko}} \bibnamefont{and} \bibinfo{author}{\bibfnamefont{S.~J.} \bibnamefont{Tyler}}, ``{The one-loop effective potential of the Wess-Zumino model revisited},'' \bibinfo{journal}{JHEP} \textbf{\bibinfo{volume}{09}}, \bibinfo{pages}{135} (\bibinfo{year}{2014}), \eprint{1407.5270}.

\bibitem[{\citenamefont{Buchbinder et~al.}(1994)\citenamefont{Buchbinder, Kuzenko, and Yarevskaya}}]{BUCHBINDER1994665}
\bibinfo{author}{\bibfnamefont{I.}~\bibnamefont{Buchbinder}}, \bibinfo{author}{\bibfnamefont{S.}~\bibnamefont{Kuzenko}}, \bibnamefont{and} \bibinfo{author}{\bibfnamefont{J.}~\bibnamefont{Yarevskaya}}, ``Supersymmetric effective potential: superfield approach,'' \bibinfo{journal}{Nuclear Physics B} \textbf{\bibinfo{volume}{411}}, \bibinfo{pages}{665} (\bibinfo{year}{1994}), ISSN \bibinfo{issn}{0550-3213}, \urlprefix\url{https://www.sciencedirect.com/science/article/pii/0550321394904669}.

\bibitem[{\citenamefont{Harlander et~al.}(2009)\citenamefont{Harlander, Mihaila, and Steinhauser}}]{Harlander:2009mn}
\bibinfo{author}{\bibfnamefont{R.~V.} \bibnamefont{Harlander}}, \bibinfo{author}{\bibfnamefont{L.}~\bibnamefont{Mihaila}}, \bibnamefont{and} \bibinfo{author}{\bibfnamefont{M.}~\bibnamefont{Steinhauser}}, ``{The SUSY-QCD beta function to three loops},'' \bibinfo{journal}{Eur. Phys. J. C} \textbf{\bibinfo{volume}{63}}, \bibinfo{pages}{383} (\bibinfo{year}{2009}), \eprint{0905.4807}.

\bibitem[{\citenamefont{Jack et~al.}(1996)\citenamefont{Jack, Jones, and North}}]{Jack:1996vg}
\bibinfo{author}{\bibfnamefont{I.}~\bibnamefont{Jack}}, \bibinfo{author}{\bibfnamefont{D.~R.~T.} \bibnamefont{Jones}}, \bibnamefont{and} \bibinfo{author}{\bibfnamefont{C.~G.} \bibnamefont{North}}, ``{N=1 supersymmetry and the three loop gauge Beta function},'' \bibinfo{journal}{Phys. Lett. B} \textbf{\bibinfo{volume}{386}}, \bibinfo{pages}{138} (\bibinfo{year}{1996}), \eprint{hep-ph/9606323}.

\bibitem[{\citenamefont{Pickering et~al.}(2001)\citenamefont{Pickering, Gracey, and Jones}}]{Pickering:2001aq}
\bibinfo{author}{\bibfnamefont{A.~G.~M.} \bibnamefont{Pickering}}, \bibinfo{author}{\bibfnamefont{J.~A.} \bibnamefont{Gracey}}, \bibnamefont{and} \bibinfo{author}{\bibfnamefont{D.~R.~T.} \bibnamefont{Jones}}, ``{Three loop gauge beta function for the most general single gauge coupling theory},'' \bibinfo{journal}{Phys. Lett. B} \textbf{\bibinfo{volume}{510}}, \bibinfo{pages}{347} (\bibinfo{year}{2001}), \bibinfo{note}{[Erratum: Phys.Lett.B 535, 377 (2002)]}, \eprint{hep-ph/0104247}.

\bibitem[{\citenamefont{Weinberg}(1979)}]{Weinberg:1978kz}
\bibinfo{author}{\bibfnamefont{S.}~\bibnamefont{Weinberg}}, ``{Phenomenological Lagrangians},'' \bibinfo{journal}{Physica A} \textbf{\bibinfo{volume}{96}}, \bibinfo{pages}{327} (\bibinfo{year}{1979}).

\bibitem[{\citenamefont{Luty}(1998)}]{Luty:1997fk}
\bibinfo{author}{\bibfnamefont{M.~A.} \bibnamefont{Luty}}, ``{Naive dimensional analysis and supersymmetry},'' \bibinfo{journal}{Phys. Rev. D} \textbf{\bibinfo{volume}{57}}, \bibinfo{pages}{1531} (\bibinfo{year}{1998}), \eprint{hep-ph/9706235}.

\bibitem[{\citenamefont{Cohen et~al.}(1997)\citenamefont{Cohen, Kaplan, and Nelson}}]{Cohen:1997rt}
\bibinfo{author}{\bibfnamefont{A.~G.} \bibnamefont{Cohen}}, \bibinfo{author}{\bibfnamefont{D.~B.} \bibnamefont{Kaplan}}, \bibnamefont{and} \bibinfo{author}{\bibfnamefont{A.~E.} \bibnamefont{Nelson}}, ``{Counting 4 pis in strongly coupled supersymmetry},'' \bibinfo{journal}{Phys. Lett. B} \textbf{\bibinfo{volume}{412}}, \bibinfo{pages}{301} (\bibinfo{year}{1997}), \eprint{hep-ph/9706275}.

\end{thebibliography}

\end{document}